\documentclass[aps,prl,twocolumn,superscriptaddress]{revtex4-1}

\usepackage[colorlinks=true, allcolors=blue]{hyperref}
\usepackage[colorinlistoftodos]{todonotes}
\usepackage{upgreek} 
\usepackage{natbib}
\usepackage{graphicx}

\begin{document}

\title{Imaging recoil ions from optical collisions between ultracold, metastable neon isotopes}

\author{B. Ohayon}
\email{ben.ohayon@mail.huji.ac.il}
\author{H. Rahangdale}
\author{J. Chocron}

\affiliation{
 Racah Institute of Physics, Hebrew University, Jerusalem 91904, Israel
}
\author{Y. Mishnayot}

 \affiliation{
 Racah Institute of Physics, Hebrew University, Jerusalem 91904, Israel
}
 \affiliation{
 Soreq Nuclear Research Center, Yavne, 81800, Israel 
}

\author{R. Kosloff}
\affiliation{
 The Institute of Chemistry, The Hebrew University of Jerusalem, Jerusalem 91904, Israel.
}
\author{O. Heber}

\affiliation{
The Weizmann Institute of Science, Rehovot 76100, Israel
}

\author{G. Ron}

 \affiliation{
 Racah Institute of Physics, Hebrew University, Jerusalem 91904, Israel
}
\date{December 2018}

\date{\today}

\begin{abstract}

We present an experimental scheme which combines the well established method of velocity-map-imaging, with a cold trapped metastable neon target. 
The device is used for obtaining the branching ratios and recoil-ion energy distributions for the penning ionization process in optical collisions of ultracold metastable neon. The potential depth of the highly excited dimer potential is extracted and compared with theoretical calculations.
The simplicity to construct, characterize, and apply such a device, makes it a unique tool for the low-energy nuclear physics community, enabling opportunities for precision measurements in nuclear decays of cold, trapped, short-lived radioactive isotopes.

\end{abstract}

\maketitle

Excited states of noble gases harbor sufficient energy to ionize most atoms and molecules upon collision. These reactive collisional processes play a crucial role in early universe chemistry \cite{2012-Astro}, have been identified as important for the evolution of planetary atmospheres \cite{2015-Vech}, and the chemistry of cold interstellar medium \cite{2018-Clouds}.

Noble gas atoms posses a long-lived \textit{metastable} state which is amenable to the methods of laser-cooling and trapping \cite{2007-Metcalf,2012-Birkl}, leading to ultracold ($<$mK), dense ($>10^{10}$/cm$^3$) samples, in which collisional processes occur predominantly in the quantum regime, where only a few partial waves contribute. However, a major difficulty arises for quantum mechanical calculations of these reaction processes, owing to the coupling of the entrance molecular channel to the ionization continuum \cite{2015-Difficult}, and the highly excited nature of the reactants \cite{2016-Shagam}. 

\begin{figure}[!htbp]
\centering
\includegraphics[width=\columnwidth,trim={0 8 0 0},clip]{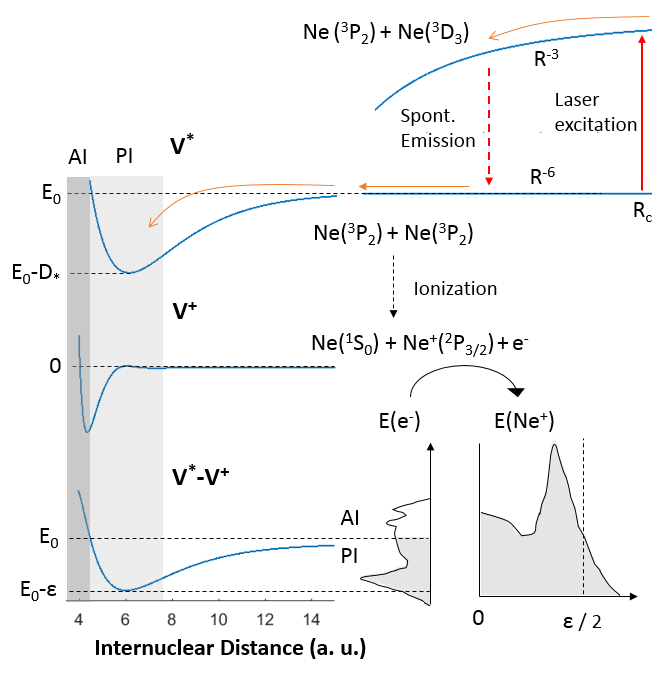}
\caption{
Schematic representation of optically assisted collisions.
Ultracold, metastable neon atoms are excited by near resonant laser radiation from a weak Van-Der-Waals attraction, to a strong dipole-dipole attraction at a large internuclear distance $R_c$, where they accelerate until emitting a photon spontaneously. 
Associative ionization (AI) occurs in the dark shaded area of the difference potential and leads to electrons with energy higher than the available energy $E_0=12$ eV.
Penning ionization (PI) occurs in the light shaded area and leads to electrons with energy lower than $E_0$. The lowest energy electrons, which correspond to the highest energy penning ions, result from ionization near the minimum of the difference potential and grant access to the potential depth $D_*\approx\epsilon$.
}
\label{fig:process}
\end{figure}

An appreciable microscopic understanding of these reactions stems from an agreement between state of the art, relativistic \textit{ab initio} potentials and ionization widths, and ample, precise experimental input. 
Experiments investigating single or dual species ultracold collisions, usually measure trap loss and ionization rates \cite{2016-Shagam, 2002-Kup, 2006-stas, 2006-ArRb, 2011-OpticalCollisions, 2016-HeRb}, and in some implementations employ mass-spectroscopic techniques for separating the reaction products \cite{1998-Mast,2008-REMPI,2013-REMPI,2019-NeAr}. Thus, theoretical calculations, which have many degrees of freedom, are calibrated by comparing to a single, global quantity such as a reaction cross-section or the branching ratio to an ionic species \cite{2018-Wasler,2019-UltracoldHe}. However, accurate determinations of the energy distributions of electrons and ions, routinely employed for molecular beams \cite{1993-Siska}, offer a high resolution window as to the involved potential energy surfaces (Fig. \ref{fig:process}).

Modern determinations of charged fragments angle-resolved energy distributions usually rely on two powerful techniques: Cold Target Recoil Ion Micrsocopy (COLTRIMS), reviewed in \cite{2000-COLTRIMS_Review}, and Velocity Map Imaging (VMI) \cite{1997-EP,2017-Parker}. The merging of these methods with laser-cooling and trapping techniques, offers substantial advantages over beam experiments and opens up new areas of research. 
Moreover, ultracold samples harbor the possibility of quantum-state preparation, and coherent control of collisional properties \cite{2006-Coherent}.
From a nuclear physics point-of-view, a cold, trapped, short-lived (typically $\tau = 0.1-100$ s) sample, constitutes the ideal target for precision Beta- \cite{2009-BehrTraps}, and Beta-delayed-neutron- \cite{2013-Delayed} decay studies.

\begin{figure}[!tp]
\includegraphics[width=1\columnwidth,trim={0 0 0 0},clip]{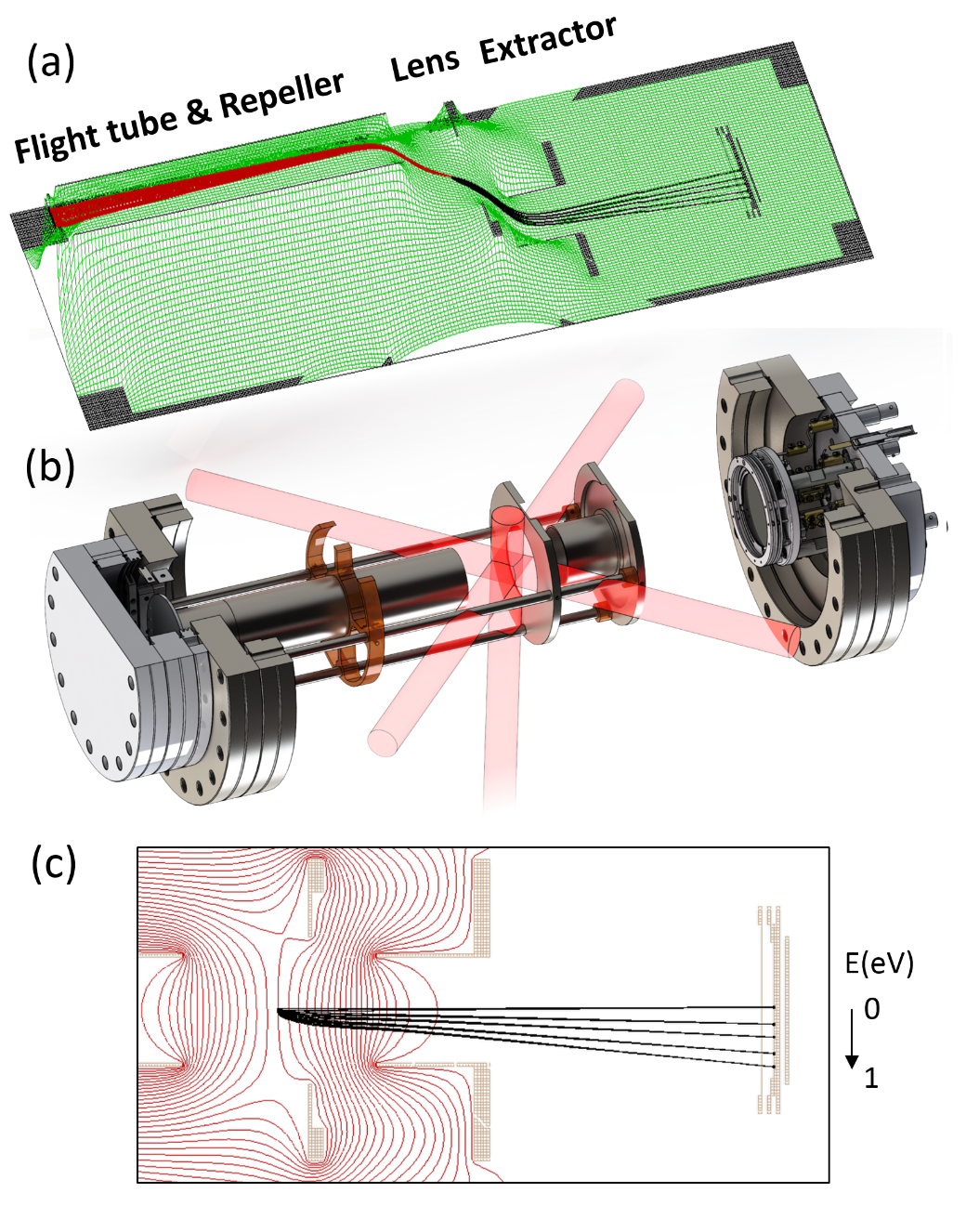}
\caption{\label{fig:CAD} The Magneto-Optical-Trap Velocity-Map-Imaging (MOT-VMI) setup.
(a) SIMION 8.1 simulation of the trajectories of charged particles emerging from the trap volume in time-of-flight mode of operation. At $4$ kV deflection voltage, all ions (red trajectories) up to $1$ eV reach the detector at the end of the flight tube, and all electrons up to $13$ eV (black trajectories) reach the position sensitive detector. Electrical potential at the center plane is portrayed as a gravitational surface.
(b) CAD drawing of the implemented device. The trapping region is located in the intersection of the laser-cooling beams.
(c) SIMION 8.1 simulation of ion trajectories in ion-imaging mode of operation. The ions emerge perpendicularly to the detector, from a $1$ mm FWHM trap volume, grouped by initial energy of $0-1$ eV, and focused to less than $200$  $\upmu$m on a position sensitive detector. Potential contours of $5$ V spacing are indicated.}
\end{figure}

Owing to the aforementioned advantages, COLTRIMS has been successfully incorporated with a Magneto-Optical-Trap (MOT) target to create the so-called MOTRIMS devices \cite{2003-MOTRIMS}. Nevertheless, the cost, complexity and the level of involvement of its operation \cite{2013-COLTRIM_VMI_Comp}, as well as the inherently low rates and efficiencies \cite{2014-Kling_Thick}, have limited the use of MOTRIMS to a few groups within the atomic and molecular physics community \cite{2008-MOTRIMS,2009-MOTIMRS}, and precluded it from being adopted by nuclear physicists.

In this letter we report a successful implementation of a simple, efficient, MOT-VMI device. Utilizing its abilities, we report precise values for penning and associative ionization (PI and AI respectively) branching ratios (BRs), as well as recoil ion energy distribution from cold, optical collisions between trapped metastable neon isotopes. From the energy distributions we extract the potential well depth of the highly excited molecular potential, and compare with \textit{ab initio} calculations and similar systems.

A traditional VMI setup consists of a repeller electrode, a gridless ring electrode acting as an electrostatic lens, and a gridless, grounded ring extractor \cite{1997-EP}.
The voltage difference between the repeller and extractor, along with the size and distance of the imaging detector, sets the scale of energies to be detected. 
The lens voltage and shape are optimized so as to cancel the effect of the finite source size. Particles with the same momentum, emerging from different locations within the target volume are imaged to the same location on the detector. Once the dynamic range is selected, the device is optimized by fine adjustment of the lens voltage.
The energy resolution in most VMI setups is determined by the effective target volume, the quality of focusing and the intrinsic resolution of the detection system. For low to medium ($10$ eV) energy charged particles, most reported values are in the range of $\Delta E/E = 1-5\%$, which is accomplished here as well.

Figure \ref{fig:CAD}, presents our adaptation of the VMI geometry to the magneto-optical-trap environment. The cooling and trapping setup includes a dynamically reconfigurable Zeeman-Slower \cite{2015-Zeeman}, and a state- and isotope-selective deflection stage in \cite{2018-Mardor}.
For merging the VMI and the MOT, we elected to use the simple, cylindrically-symmetric, three-electrode arrangement, where we merged the repeller with a flight-tube for coincidence time-of-flight (TOF) detection. The radii and positions of various elements are optimized within the geometrical constraints of the MOT using the simulation package SIMION 8.1, to maximize the focusing capabilities for a variety of modes-of-operation.

The internal energy of cold, trapped metastable neon, $E^*=17$ eV, exceeds the ionization potential $E^+$ of most atoms and molecules. The ionization reaction between a metastable noble atom Rg$^*$ and a molecule AB may be described schematically as:
\[
\mathrm{Rg^*+AB} \leftrightarrow  \mathrm{[RgAB]^*} \rightarrow \mathrm{[RgAB]}^++e^-,
\]
where the excited molecule [RgAB]$^*$ autoionizes instantly to form a nascent ionic complex. The emitted electron takes most of the available energy $E_0=E^*-E^+$, while a portion of it is deposited in ro-vibrational levels of the nascent ion.
The resulting ionic complex may remain intact, in which case the entire process is called associative ionization (AI),
or dissociate to a number of possible channels \cite{1976-Cehniionization}:

\begin{tabular}{lll}
\\
  $ \mathrm{[RgAB]}^+  $  &  $ \rightarrow \;\;\mathrm{Rg+AB^+}  $   &   ~  (PI)   \\
               &  $\rightarrow \;\;\mathrm{RgA + B^+  }$    &  ~    (RI)         \\
               &  $\rightarrow \;\;\mathrm{Rg+A + B^+  }$   &  ~   (DI).        \\\\
  \end{tabular}%
  
\noindent 
These are labeled penning ionization (PI), rearrangement ionization (RI), and dissociation ionization (DI). Considering both cold collisions within the trap (intra-trap) and with thermal background gasses (inter-trap), all of the aforementioned reactions are present in our system.
\begin{figure}[!tp]
\centering
\includegraphics[width=\columnwidth,trim={18 8 7 10},clip]{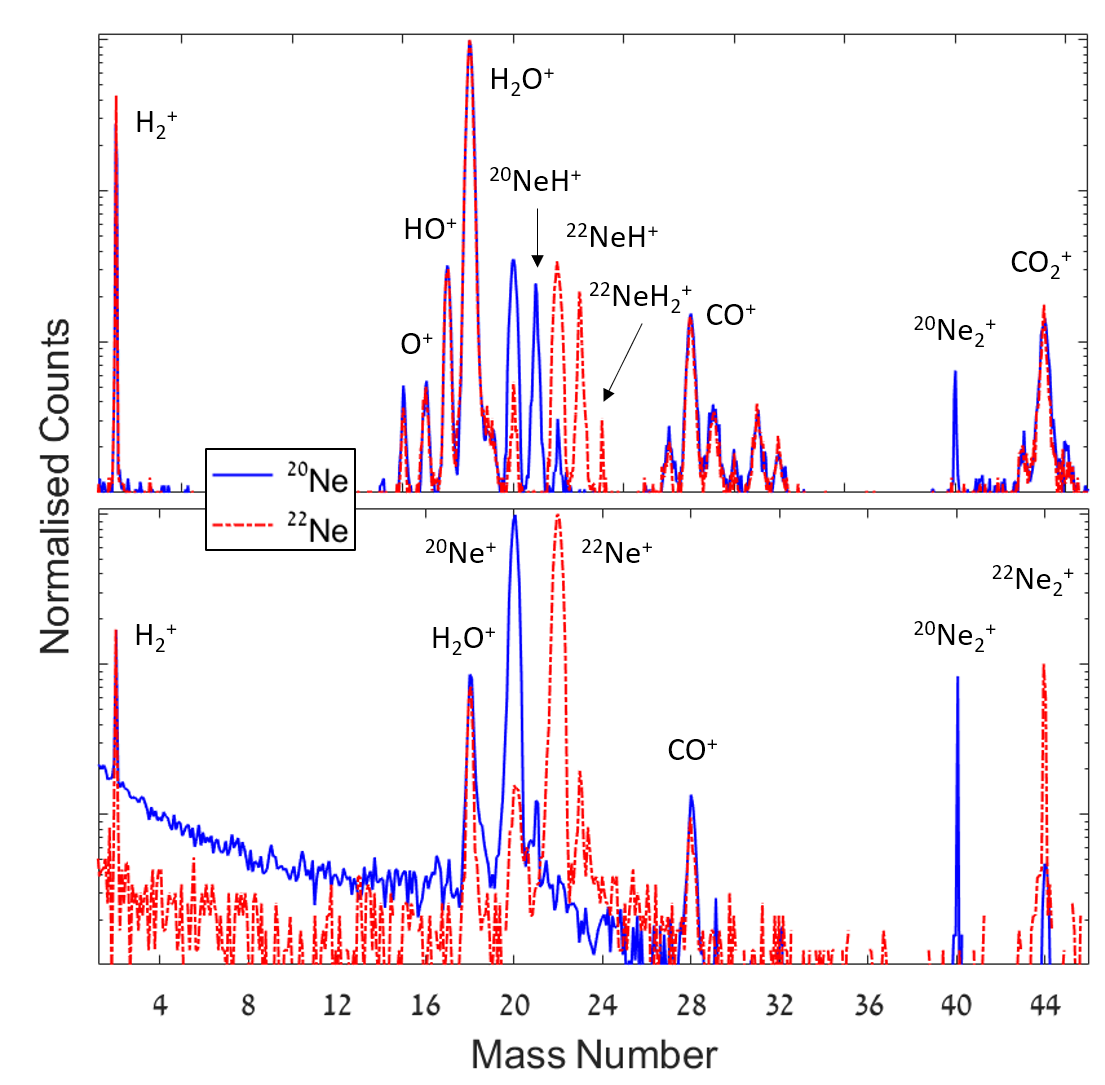}
\caption{
Mass spectra of recoiling ions from collisions with trapped neon isotopes. Counts are presented in logarithmic scale and normalised to the highest peak. The top spectrum was taken at low trap density and is dominated by thermal inter-trap collisions. The bottom spectrum was taken at high trap density and ultra-high-vacuum, and is dominated by ultracold, intra-trap collisions.
}
\label{fig:c}
\end{figure}
Figure \ref{fig:c} presents the mass-calibrated time-of-flight spectrum of recoil ions detected when trapping $^{20}$Ne$^*$ or $^{22}$Ne$^*$. 

\begin{table}[!bp]
  \centering
  \caption{Branching ratios for ionizing collisions between Ne$^*$ and H$_2$O molecules. The neon isotope mass number, and average collision energy are tabulated.}
 \begin{ruledtabular} 
    \begin{tabular}{llllllllll}
 \textbf{Mass} &  \textbf{E$_{avg}$}    & \textbf{H$_2$O$^+$}  & \textbf{HO$^+$}   & \textbf{O$^+$} & Ref. \\
\textbf{number}& meV  &   \%     & \%   &  \%     \\
\cline{1-6}
20   & 17  & 96.2(5)   & 3.3(5)      & 0.46(3)  & This Work  \\
 22  & 18   & 96.8(5)   & 2.8(5)     & 0.45(10)  & This Work \\
20+22   & 55   & 96.2(9.6)  & 3.2(5)   & 0.6(1)  &\cite{2015-Vech, 2016-NeH2O}  \\
20+22  & 70   & 96.7        & 2.9      & 0.4 &  \cite{2012-H2O}\\

\end{tabular}%
\end{ruledtabular}
\label{tab:H2O_1}%
\end{table}%

At moderate vacuum conditions, and low trap density, water peaks are prominent, due to the large
ionization cross section of H$_2$O-Ne$^*$, resulting from their strong
attraction \cite{2013-Stero}. We checked that branching ratios (BR) for ions
resulting from Ne$^*$-H$_2$O collisions are stable with changing laser power,
detuning, magnetic field strength, and by alternating between $^{20,22}$Ne. These
are reported in table \ref{tab:H2O_1}, and are in superb agreement with \cite{2012-H2O}, who utilized a crossed-beam technique. In accordance with \cite{2012-H2O}, no evidence (BR $<0.1\%$) was found for AI, indicated by the lack of NeH$_2$O$^+$
ions.

The trap density is many orders of magnitude higher than the Ne$^*$ background and so neon ions and dimers result solely from intra-trap collisions.
In the presence of a strong, near resonant, light field, half of the atoms are in the $^3$D$_3$ excited state, where they experience a strong attraction to the metastable $^3$P$_2$ state, due to the resonant dipole-dipole interaction between them. In these optically-assisted collisions, portrayed in Fig. \ref{fig:process}, the two atoms are accelerated toward small internuclear distances where the probability of ionization is significantly larger. This process increases the collision rates by two orders of magnitude \cite{2011-OpticalCollisions}.

From mass spectra similar to Fig. \ref{fig:c}, we find that the presence of near-resonant laser light, increased the formation of dimers by at least a factor of two, providing strong evidence for photoassociation in cold neon collisions (see the Supplemental Material \cite{Sup}, which includes refs \cite{1997-MCPsup,2019-ISsup,2015-RFsup}, for a discussion on the extraction of branching ratios). This enhancement is qualitatively understood from the process portrayed in Fig. \ref{fig:process}, where the accelerated dimer at the potential $V^*(R)$ has a larger probability of reaching a small internuclear distance of roughly $4$ a.u. where the ionic potentials $V^+(R)$ become strongly attractive, leading to positive difference potentials which promote association.
The branching ratios for the creation of neon dimer ions are presented in table \ref{tab:BR} along with collision rates from the literature, and compared with results for trapped metastable helium. Previous investigations into optical collisions in neon traps, by \cite{2008-Drunen} and \cite{2011-OpticalCollisions}, did not detect PI and AI separately.

\begin{table}[!bp]
  \centering
  \caption{Branching ratios and rates for ionizing collisions between trapped metastable atoms. 
  $K_{gg}$ is the two-body total ionization rate between ground state atoms.
  $K_{ge}$ is taken at near resonant light conditions, where reactive collisions occur predominantly between ground and excited state atoms.}
  \begin{ruledtabular}
    \begin{tabular}{llll}
        & $K_{gg}$ cm$^3$/s     & $K_{ge}$ cm$^3$/s & Ref. \\
\cline{1-4}        
Ne$_2$$^+$+Ne$^+$   &   $4(1)\times10^{-10}$      & $2.0(3)\times10^{-8}$& \cite{2011-OpticalCollisions,2002-Kup} 
\\
Ne$_2$$^+$/Ne$^+$ & $<2.5\%$ & $5.1(1)\%$ & This Work
\\\\

He$_2$$^+$+He$^+$ & $1.0(2)\times10^{-10}$&   $1.3(3)\times10^{-8}$ &\cite{1999-Tol,2006-stas} 
\\
He$_2$$^+$/He$^+$ & $3.0(3)\%$  & $16(2)\%$& \cite{1998-Mast} \\  
  \end{tabular}%

  \end{ruledtabular}
  \label{tab:BR}%
\end{table}%

\begin{figure}[!htbp]
\centering
\includegraphics[width=\columnwidth,trim={5 8 5 8},clip]{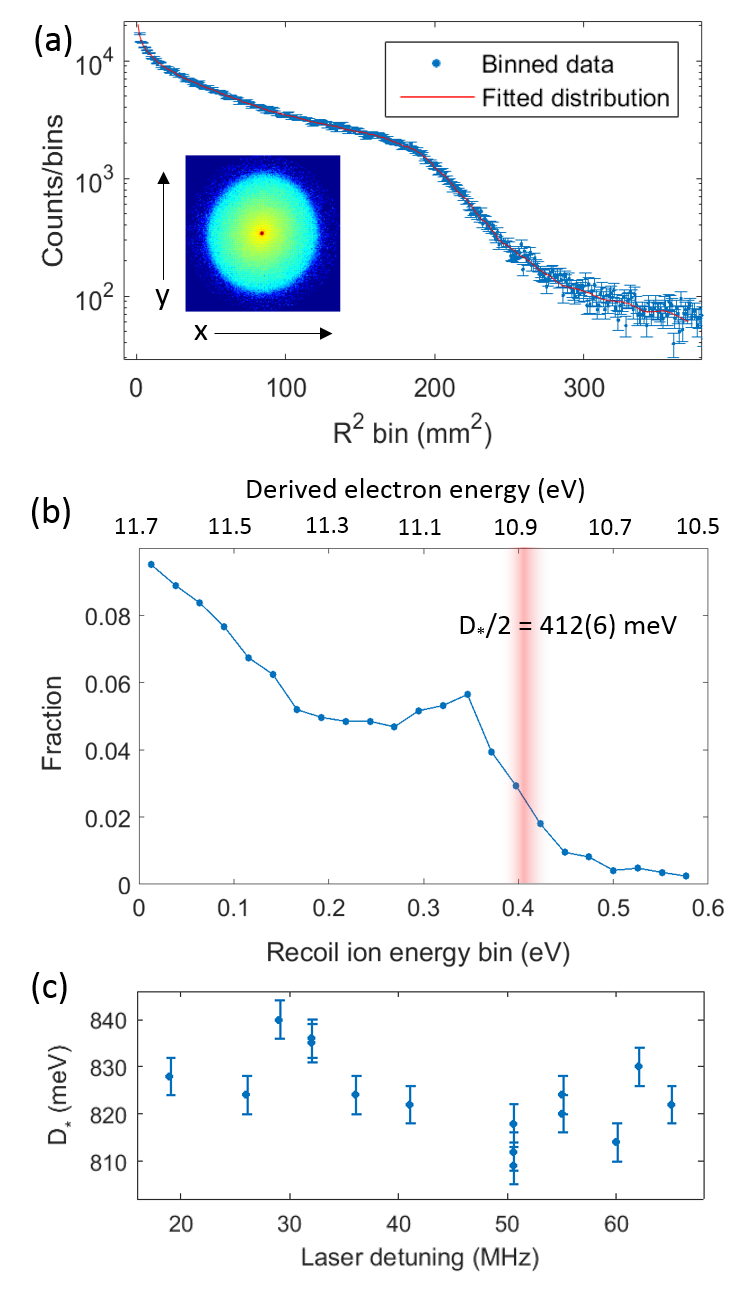}
\caption{
(a) Distribution of squared hit radii from 2D image on the detector plane (inset). The solid line indicates the best fit from the inversion process.
(b) Energy spectrum of Ne$^+$ ions emerging from the (Ne$_2$)* collision complex, as extracted from fitting the radial hit distribution (see Supplemental Material, which includes Ref. \cite{2003-imagingSup}, for further discussion on the inversion method). Electron energies derived from equation \ref{eq.E} are indicated, as well as the resulting well depth $D_*$.
(c) Determined well depth for various red detuning of the trapping laser.}
\label{fig:VMI}
\end{figure}

The MOT-VMI enables high-resolution detection of the recoil energy distribution of penning ions from ultra-cold collisions. An example of which, taken without coincidence-detection, with a $100$ times less water background, and $\times200$ denser trap, as compared to the conditions of Fig. \ref{fig:c} top, is shown in Fig. \ref{fig:VMI}.  
Since the reactants are cold, the penning ion kinetic energy $E_i$, is equal to half of the missing energy of the penning electrons \cite{2002-RIMSHe} 
\begin{equation}\label{eq.E}
E_i=(E_0-E_e)/2.
\end{equation}
Thus, it offers a high resolution window, free of the $E_0=12$ eV offset, into the quantum-governed dynamics of the penning ionization process. As shown in figure \ref{fig:process}, the lowest energy electrons, which correspond to the highest energy ions, result from ionization near the minimum $E_0-\epsilon$ of the difference potential $V^*-V^+$. 
Following the recipe of \cite{1970-Miller,1974-Hotop} usually applied to electron spectra, and utilizing Eq. \ref{eq.E}, $\epsilon/2$ may be extracted in a straightforward manner from the $44\%$ edge at the high energy side of the penning ion distribution (Fig. \ref{fig:VMI}b).
To extract information on $V^*$, we utilize the consistency between various estimations of the location of minima of $V^*$, at $R_\mathrm{min}=6.0-6.5$ a.u. \cite{1998-Doery, 2000-Tupitsyn}, and the observation that the well-known ionic potentials do not vary substantially from zero around $R_\mathrm{min}$ \cite{1984-Photoionzation,2002-Microwave}. Thus the well-depth of lowest potential may be evaluated directly from
\begin{equation}
D_* = \epsilon-V^+(R_\mathrm{min}) \approx \epsilon,
\end{equation}

The results of various determinations of $D_*$ are presented in Fig. \ref{fig:VMI}c, where they are found to not vary substantially with laser detuning. This observation further indicates that most optical ionization collisions in neon occur after the spontaneous emission of a photon, in contrast with the case for He$^*$ \cite{1998-Mast}, which has a longer lived excited state and a smaller mass (see Supplemental Material \cite{Sup}, which includes Refs. \cite{1988-DoerySup,1989-GPsup}, for further discussion on the probability for spontaneous emission).

We estimate the uncertainty of the most repulsive $V^+(R_\mathrm{min})$ as $10$ meV resulting from the uncertainty in $R_\mathrm{min}$ and conclude that the deepest potential depth for Ne$^*$($^3$P$_2$)-Ne$^*$($^3$P$_2$) is $D_*=824(22)$ meV.
Kotochigova \textit{et. al.} \cite{2000-Tupitsyn}, utilized a nonrelativistic multiconfiguration valence-bond method to calculate these potentials at short range \textit{ab initio}. They report that the deepest attractive potential, labeled $^1\Delta_\sigma$, has a well-depth of $D_*=437$ meV, in strong disagreement with our findings. 
This discrepancy is to some extent resolved when considering approximate short-range potentials based on Na$_2$ developed by \cite{1998-Doery}, where the inclusion of spin-orbit interactions increased the well-depths by roughly $300$ meV.
Our result is compared with similar collisional systems in table \ref{tab:Well}.

To conclude, we successfully implemented a simple, versatile, MOT-VMI device, and demonstrated a few of its applications by conducting precise measurements of branching ratios and energy spectra of recoil ions emerging from inter- and intra- trap collisions. The branching ratios for ionizing process in metastable neon colliding with water molecules are in superb agreement with those measured in crossed-beam experiments and may be beneficial for advancing the understanding of the penning ionization processes in planetary atmospheres \cite{2015-Vech}. 
Through the increase in the branching ratio for associative ionization in the presence of the trapping laser, we find long sought evidence for photoassociation processes in noble gasses other than helium.
Utilizing the imaging capabilities and a fast and simple inversion scheme, we obtain the energy distribution of recoil neon ions from cold optical collisions within the trap. The well-depth of the lowest, highly excited molecular potential, is extracted, and disagrees with nonrelativistic \textit{ab initio} calculations, demonstrating the dramatic effect of spin-orbit coupling, and the necessity of including relativistic effects in \textit{ab initio} calculations of highly excited molecular potentials.
The simplicity of construction, operation, and data analysis of the MOT-VMI makes it a compelling new tool for investigations of reactive processes in ultracold chemistry, and coherent control of the outcome of ionizing collisions \cite{2006-Coherent, 2015-Nanosecond, 2018-Coherent}.

Our entire system is located above the beamline of the Soreq Applied Research Accelerator (SARAF) \cite{2018-Mardor}, and we intend to utilize the MOT-VMI for precision measurements of recoil ions from the beta-decay of short-lived neon isotopes, in search of new physics in the weak sector of the standard model \cite{2018-Weak}.

\begin{table}[htbp]
  \centering
  \caption{Experimental well depths $D_*$, in meV, of the lowest diatomic potentials in Ne$^*$-Ne$^*$ collisions, and similar systems.}
  \begin{ruledtabular}
    \begin{tabular}{p{1.7cm}ccccc}
& He$^*$($2s$$^3$S) & Ne$^*$($3s$$^3$P$_2$)  & Li($2s$$^2$S)     &  Na($3s$$^2$S)   
 \\

\\He$^*$($2s$$^3$S)  & 850(90)$^a$  & 500(100)$^b$   & 868(20)$^c$ & 740(25)$^c$ 
\\Ne$^*$($3s$$^3$P$_2$) & 500(100)$^b$ & 824(22)$^d$  & 798(30)$^e$ & 678(18)$^f$  \\ Ar$^*$($4s$$^3$P$_2$) &  & 300(50)$^g$ &    &602(23)$^f$  \\\\

\begin{minipage}[t]{0.5\columnwidth}
$^a$ Estimation based on \cite{2002-RIMSHe}
  \end{minipage}\tabularnewline
~  $^b$ Ref. \cite{1980-HeNe}&
  $^c$ Ref. \cite{1998-Hotop}&
  $^d$ This Work \\
~ $^e$ Ref. \cite{1986-alkali} &
 $^f$ Ref. \cite{1990-RgNa}&
  $^g$ Ref. \cite{1980-NeAr}
 \end{tabular}%
\end{ruledtabular}
 
\label{tab:Well}%
\end{table}%

The work presented here is supported by grants from the Pazy Foundation (Israel), Israel Science Foundation (grants no. 139/15 and 1446/16), and the European Research Council (grant no. 714118 TRAPLAB). BO is supported by the Ministry of Science and Technology, under the Eshkol Fellowship.

\section{Supplemental Material}

\subsection{Probability of spontaneous emission}

The atomic pair is excited from a weak Van der Waals attractive potential to a strong dipole-dipole potential $\pm C_3/r^3$. Out of the $40$ excited molecular potentials, the strongest attractive potentials have $C_3$ coefficients of up to $-10$ a.u. \cite{1997-Verhhar}.

Within the Gallagher-Pritchard model for optically assisted collisions \cite{1989-GP}, the maximal semi-classical excitation rate of the pair, is attained at the Condon distance $R_c=(C_3/\hbar\delta)^{1/3}$, corresponding to $R_c=1000-1500$ a.u. for red detunings of $\delta=(20-60)\times2\pi$ MHz. The time it takes to reach a small (few a.u.) interatomic distance, where ionization is probable, is $\tau_p=\sqrt{\frac{\mu R_c^5}{2 C_3}}=40$ ns.
The probability for molecular spontaneous emission, with rate $\Gamma_M=16\times2\pi$ MHz, before the ionization process occurs, is $1-\exp(-\Gamma_M\tau_p)=99\%$ \cite{1989-GP}.

We conclude that at the small detunings considered in this work, most collisions occur at the ground Ne*-Ne* potential, following light-induced acceleration. A future direction for this work is the introduction another intense laser beam at large red detuning ($\delta<-1$ GHz) to induce ionizing collisions in the molecular laser-excited states.

\subsection{Extraction of branching ratios}

The branching ratios for Ne$_2^+$/Ne$^+$ are extracted from calibrated mass spectra in efficient time-of-flight (TOF) mode. Ions are accelerated in the flight tube by keeping it and the micro-channel-plate (MCP) detector top plate at $-4$ kV, ensuring similar quantum efficiency for a variety of ionic masses \cite{1997-MCP}. Electrons are deflected by the flight tube and are efficienctly detected in coincidence on a grounded position sensitive MCP, in which their hit position is monitored.
For each measurement sequence, we alternated between trapping of $^{20}$Ne and $^{22}$Ne by offsetting the trapping laser frequency by their isotope shift of $1.6260$ GHz \cite{2019-IS}. A $45$ degree isotopic-selective deflection stage ensures that only metastable states of the selected isotope reach the trapping volume, and that the trapped cloud is not ionized by VUV radiation emanating from the metastable-source \cite{2015-Source}.

\begin{figure}[!htbp]
\centering
\includegraphics[width=0.8\columnwidth,trim={2 15 2 2},clip]{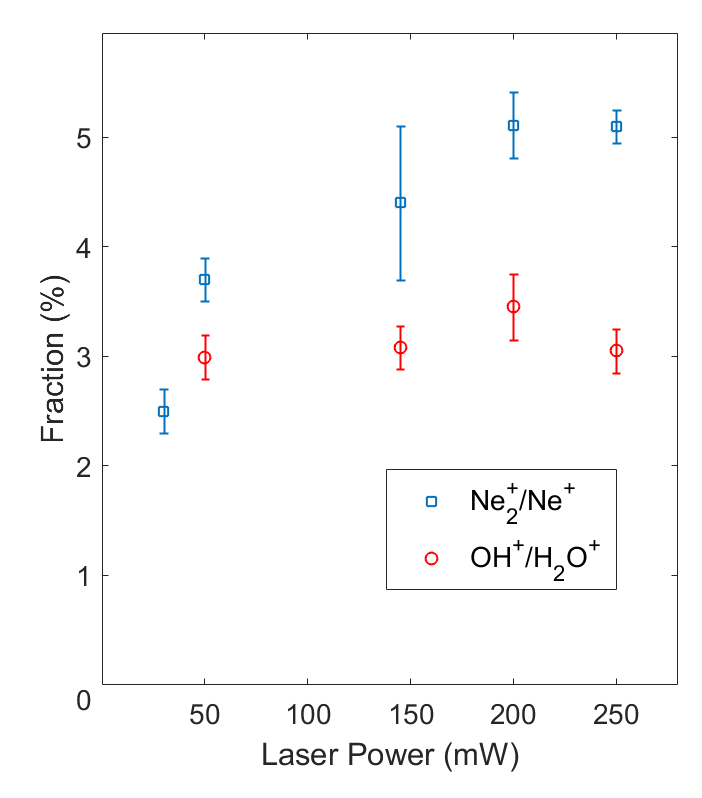}
\caption{
Ne$_2^+$/Ne$^+$ branching ratio (blue squares) and OH$^+$/H$_2$O$^+$ branching ratio (red circles), as a function of trapping laser power
}
\label{fig:BR}
\end{figure}

Fig. \ref{fig:BR} presents the  branching ratios (BR) extracted from mass spectra, as a function of trapping laser power. Uncertainties are dominated by counting statistics due to the small fraction of Ne$_2^+$, as well as employing different models for the background counts. The branching ratio displays a pronounced enhancement when increasing laser power. To test that this enhancement is not an artifact of the change in trap density or position, we extract for each spectra the OH$^+$/H$_2$O$^+$ branching ratio as well, which has similar uncertainty, and does not show such an enhancement. For the minimal laser power, the use of an ion pump to remove the water background precluded the measurement of the OH$^+$/H$_2$O$^+$ BR.

We find that the Ne$^+$/H$_2$O ratio increases by roughly a factor $60$ between minimal and maximal trapping laser power, while keeping the neon flux and vacuum conditions constant, indicating a large enhancement of the intratrap ion yield in the presence of laser light. This increase is in accordance with the literature values cited in table II of the main text.

A future direction of this experiment is an implementation of fast-switching of the trapping laser, in coincidence with the detection system, to extract the BR for collisions in the dark.


\subsection{Inversion of VMI image}

In most cases where a VMI image requires inversion, one uses one of the many fast, efficient and well tested Abel-inversion codes \cite{2003-Imaging}. Nevertheless, we opted to use direct propagation relying on a full Simion 8.1 simulation, taking into account the trap size, magnetic-field, and the shape of the electrodes, detectors, and various ground planes.

For each experimental configuration, a few $10^7$ ions are sampled from a uniform energy and emission angle distributions, and a $1-3$ mm FWHM 3D Gaussian trap volume depending on the laser detuning. The ion trajectories are simulated in the electromagnetic potential generated by the electrodes and magnetic coils, and their hit position $(x,y)$ on the detector, as well as the starting kinetic energy, are recorded. 

\begin{figure}[!bp]
\centering
\includegraphics[width=1\columnwidth,trim={2 5 2 2},clip]{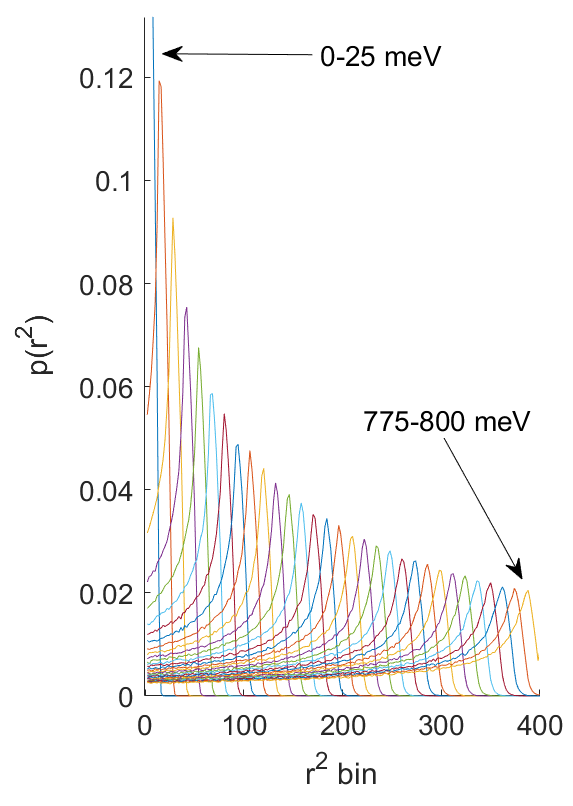}
\caption{Normalised basis functions for inversion. Each line is a simulated histogram of squared hit radii corresponding to a different energy group.
}
\label{fig:Inv}
\end{figure}

Due to the cylindrical symmetry of the electrode system, and the spherically symmetric collision process in a magneto-optical-trap, we consider the squared hit radius $r^2=(x-x_0)^2+(y-y_0)^2$ 
of the ions, where $(x_0,y_0)$ is the detector center. The ions' $r^2$ are then grouped by starting energy with typically $N=40-60$ energy bins. For each image, the simulated and measured $r^2$ hits are grouped in $400$  bins with identical bin edges.

The matter of extracting ion energy distributions is now reduced to the determination of $N$ parameters $a_i$ which correspond to the scaling of a specific simulated energy group $r^2$ histogram, and a background parameter $b$ which is independent of hit position. An example of these histograms is shown in Fig. \ref{fig:Inv}.

Fitting to the experimental data is accomplished through a standard minimization of
\[
\chi^2(a_i,b)=\sum_{i}\frac{(n(i)-\sum_{j}a_in_s(i,j)+b)^2}{n(i)},
\]
with $n(i)$ the measured number of hits in $r^2$ bin number $i$, and $n_s(i,j)$ the simulated number of hits in $r^2$ bin number $i$ and energy group $j$. The uncertainty is dominated by the statistics of roughly $5\times10^5$ measured events.

\bibliography{references}

\begin{thebibliography}{66}%
\makeatletter
\providecommand \@ifxundefined [1]{%
 \@ifx{#1\undefined}
}%
\providecommand \@ifnum [1]{%
 \ifnum #1\expandafter \@firstoftwo
 \else \expandafter \@secondoftwo
 \fi
}%
\providecommand \@ifx [1]{%
 \ifx #1\expandafter \@firstoftwo
 \else \expandafter \@secondoftwo
 \fi
}%
\providecommand \natexlab [1]{#1}%
\providecommand \enquote  [1]{``#1''}%
\providecommand \bibnamefont  [1]{#1}%
\providecommand \bibfnamefont [1]{#1}%
\providecommand \citenamefont [1]{#1}%
\providecommand \href@noop [0]{\@secondoftwo}%
\providecommand \href [0]{\begingroup \@sanitize@url \@href}%
\providecommand \@href[1]{\@@startlink{#1}\@@href}%
\providecommand \@@href[1]{\endgroup#1\@@endlink}%
\providecommand \@sanitize@url [0]{\catcode `\\12\catcode `\$12\catcode
  `\&12\catcode `\#12\catcode `\^12\catcode `\_12\catcode `\%12\relax}%
\providecommand \@@startlink[1]{}%
\providecommand \@@endlink[0]{}%
\providecommand \url  [0]{\begingroup\@sanitize@url \@url }%
\providecommand \@url [1]{\endgroup\@href {#1}{\urlprefix }}%
\providecommand \urlprefix  [0]{URL }%
\providecommand \Eprint [0]{\href }%
\providecommand \doibase [0]{http://dx.doi.org/}%
\providecommand \selectlanguage [0]{\@gobble}%
\providecommand \bibinfo  [0]{\@secondoftwo}%
\providecommand \bibfield  [0]{\@secondoftwo}%
\providecommand \translation [1]{[#1]}%
\providecommand \BibitemOpen [0]{}%
\providecommand \bibitemStop [0]{}%
\providecommand \bibitemNoStop [0]{.\EOS\space}%
\providecommand \EOS [0]{\spacefactor3000\relax}%
\providecommand \BibitemShut  [1]{\csname bibitem#1\endcsname}%
\let\auto@bib@innerbib\@empty
\bibitem [{\citenamefont {Gerlich}\ \emph {et~al.}(2012)\citenamefont
  {Gerlich}, \citenamefont {Jusko}, \citenamefont {Rou{\v{c}}ka}, \citenamefont
  {Zymak}, \citenamefont {Pla{\v{s}}il},\ and\ \citenamefont
  {Glos{\'\i}k}}]{2012-Astro}%
  \BibitemOpen
  \bibfield  {author} {\bibinfo {author} {\bibfnamefont {D.}~\bibnamefont
  {Gerlich}}, \bibinfo {author} {\bibfnamefont {P.}~\bibnamefont {Jusko}},
  \bibinfo {author} {\bibfnamefont {{\v{S}}.}~\bibnamefont {Rou{\v{c}}ka}},
  \bibinfo {author} {\bibfnamefont {I.}~\bibnamefont {Zymak}}, \bibinfo
  {author} {\bibfnamefont {R.}~\bibnamefont {Pla{\v{s}}il}}, \ and\ \bibinfo
  {author} {\bibfnamefont {J.}~\bibnamefont {Glos{\'\i}k}},\ }\href@noop {}
  {\bibfield  {journal} {\bibinfo  {journal} {The Astrophysical Journal}\
  }\textbf {\bibinfo {volume} {749}},\ \bibinfo {pages} {22} (\bibinfo {year}
  {2012})}\BibitemShut {NoStop}%
\bibitem [{\citenamefont {Falcinelli}\ \emph {et~al.}(2015)\citenamefont
  {Falcinelli}, \citenamefont {Pirani},\ and\ \citenamefont
  {Vecchiocattivi}}]{2015-Vech}%
  \BibitemOpen
  \bibfield  {author} {\bibinfo {author} {\bibfnamefont {S.}~\bibnamefont
  {Falcinelli}}, \bibinfo {author} {\bibfnamefont {F.}~\bibnamefont {Pirani}},
  \ and\ \bibinfo {author} {\bibfnamefont {F.}~\bibnamefont {Vecchiocattivi}},\
  }\href@noop {} {\bibfield  {journal} {\bibinfo  {journal} {Atmosphere}\
  }\textbf {\bibinfo {volume} {6}},\ \bibinfo {pages} {299} (\bibinfo {year}
  {2015})}\BibitemShut {NoStop}%
\bibitem [{\citenamefont {van~der Poel}\ and\ \citenamefont
  {Bethlem}(2018)}]{2018-Clouds}%
  \BibitemOpen
  \bibfield  {author} {\bibinfo {author} {\bibfnamefont {A.~P.}\ \bibnamefont
  {van~der Poel}}\ and\ \bibinfo {author} {\bibfnamefont {H.~L.}\ \bibnamefont
  {Bethlem}},\ }\href@noop {} {\bibfield  {journal} {\bibinfo  {journal} {EPJ
  Techniques and Instrumentation}\ }\textbf {\bibinfo {volume} {5}},\ \bibinfo
  {pages} {6} (\bibinfo {year} {2018})}\BibitemShut {NoStop}%
\bibitem [{\citenamefont {Metcalf}\ and\ \citenamefont {Van~der
  Straten}(2007)}]{2007-Metcalf}%
  \BibitemOpen
  \bibfield  {author} {\bibinfo {author} {\bibfnamefont {H.~J.}\ \bibnamefont
  {Metcalf}}\ and\ \bibinfo {author} {\bibfnamefont {P.}~\bibnamefont {Van~der
  Straten}},\ }\href@noop {} {\bibfield  {journal} {\bibinfo  {journal} {The
  Optics Encyclopedia: Basic Foundations and Practical Applications}\ }
  (\bibinfo {year} {2007})}\BibitemShut {NoStop}%
\bibitem [{\citenamefont {Vassen}\ \emph {et~al.}(2012)\citenamefont {Vassen},
  \citenamefont {Cohen-Tannoudji}, \citenamefont {Leduc}, \citenamefont
  {Boiron}, \citenamefont {Westbrook}, \citenamefont {Truscott}, \citenamefont
  {Baldwin}, \citenamefont {Birkl}, \citenamefont {Cancio},\ and\ \citenamefont
  {Trippenbach}}]{2012-Birkl}%
  \BibitemOpen
  \bibfield  {author} {\bibinfo {author} {\bibfnamefont {W.}~\bibnamefont
  {Vassen}}, \bibinfo {author} {\bibfnamefont {C.}~\bibnamefont
  {Cohen-Tannoudji}}, \bibinfo {author} {\bibfnamefont {M.}~\bibnamefont
  {Leduc}}, \bibinfo {author} {\bibfnamefont {D.}~\bibnamefont {Boiron}},
  \bibinfo {author} {\bibfnamefont {C.~I.}\ \bibnamefont {Westbrook}}, \bibinfo
  {author} {\bibfnamefont {A.}~\bibnamefont {Truscott}}, \bibinfo {author}
  {\bibfnamefont {K.}~\bibnamefont {Baldwin}}, \bibinfo {author} {\bibfnamefont
  {G.}~\bibnamefont {Birkl}}, \bibinfo {author} {\bibfnamefont
  {P.}~\bibnamefont {Cancio}}, \ and\ \bibinfo {author} {\bibfnamefont
  {M.}~\bibnamefont {Trippenbach}},\ }\href@noop {} {\bibfield  {journal}
  {\bibinfo  {journal} {Reviews of Modern Physics}\ }\textbf {\bibinfo {volume}
  {84}},\ \bibinfo {pages} {175} (\bibinfo {year} {2012})}\BibitemShut
  {NoStop}%
\bibitem [{\citenamefont {Hauser}\ \emph {et~al.}(2015)\citenamefont {Hauser},
  \citenamefont {Lee}, \citenamefont {Carelli}, \citenamefont {Spieler},
  \citenamefont {Lakhmanskaya}, \citenamefont {Endres}, \citenamefont {Kumar},
  \citenamefont {Gianturco},\ and\ \citenamefont {Wester}}]{2015-Difficult}%
  \BibitemOpen
  \bibfield  {author} {\bibinfo {author} {\bibfnamefont {D.}~\bibnamefont
  {Hauser}}, \bibinfo {author} {\bibfnamefont {S.}~\bibnamefont {Lee}},
  \bibinfo {author} {\bibfnamefont {F.}~\bibnamefont {Carelli}}, \bibinfo
  {author} {\bibfnamefont {S.}~\bibnamefont {Spieler}}, \bibinfo {author}
  {\bibfnamefont {O.}~\bibnamefont {Lakhmanskaya}}, \bibinfo {author}
  {\bibfnamefont {E.~S.}\ \bibnamefont {Endres}}, \bibinfo {author}
  {\bibfnamefont {S.~S.}\ \bibnamefont {Kumar}}, \bibinfo {author}
  {\bibfnamefont {F.}~\bibnamefont {Gianturco}}, \ and\ \bibinfo {author}
  {\bibfnamefont {R.}~\bibnamefont {Wester}},\ }\href@noop {} {\bibfield
  {journal} {\bibinfo  {journal} {Nature Physics}\ }\textbf {\bibinfo {volume}
  {11}},\ \bibinfo {pages} {467} (\bibinfo {year} {2015})}\BibitemShut
  {NoStop}%
\bibitem [{\citenamefont {Skomorowski}\ \emph {et~al.}(2016)\citenamefont
  {Skomorowski}, \citenamefont {Shagam}, \citenamefont {Narevicius},\ and\
  \citenamefont {Koch}}]{2016-Shagam}%
  \BibitemOpen
  \bibfield  {author} {\bibinfo {author} {\bibfnamefont {W.}~\bibnamefont
  {Skomorowski}}, \bibinfo {author} {\bibfnamefont {Y.}~\bibnamefont {Shagam}},
  \bibinfo {author} {\bibfnamefont {E.}~\bibnamefont {Narevicius}}, \ and\
  \bibinfo {author} {\bibfnamefont {C.~P.}\ \bibnamefont {Koch}},\ }\href@noop
  {} {\bibfield  {journal} {\bibinfo  {journal} {The Journal of Physical
  Chemistry A}\ }\textbf {\bibinfo {volume} {120}},\ \bibinfo {pages} {3309}
  (\bibinfo {year} {2016})}\BibitemShut {NoStop}%
\bibitem [{\citenamefont {Kuppens}\ \emph {et~al.}(2002)\citenamefont
  {Kuppens}, \citenamefont {Tempelaars}, \citenamefont {Mogendorff},
  \citenamefont {Claessens}, \citenamefont {Beijerinck},\ and\ \citenamefont
  {Vredenbregt}}]{2002-Kup}%
  \BibitemOpen
  \bibfield  {author} {\bibinfo {author} {\bibfnamefont {S.~J.~M.}\
  \bibnamefont {Kuppens}}, \bibinfo {author} {\bibfnamefont {J.~G.~C.}\
  \bibnamefont {Tempelaars}}, \bibinfo {author} {\bibfnamefont {V.~P.}\
  \bibnamefont {Mogendorff}}, \bibinfo {author} {\bibfnamefont {B.~J.}\
  \bibnamefont {Claessens}}, \bibinfo {author} {\bibfnamefont {H.~C.~W.}\
  \bibnamefont {Beijerinck}}, \ and\ \bibinfo {author} {\bibfnamefont
  {E.~J.~D.}\ \bibnamefont {Vredenbregt}},\ }\href@noop {} {\bibfield
  {journal} {\bibinfo  {journal} {Physical Review A}\ }\textbf {\bibinfo
  {volume} {65}},\ \bibinfo {pages} {023410} (\bibinfo {year}
  {2002})}\BibitemShut {NoStop}%
\bibitem [{\citenamefont {Stas}\ \emph {et~al.}(2006)\citenamefont {Stas},
  \citenamefont {McNamara}, \citenamefont {Hogervorst},\ and\ \citenamefont
  {Vassen}}]{2006-stas}%
  \BibitemOpen
  \bibfield  {author} {\bibinfo {author} {\bibfnamefont {R.~J.~W.}\
  \bibnamefont {Stas}}, \bibinfo {author} {\bibfnamefont {J.~M.}\ \bibnamefont
  {McNamara}}, \bibinfo {author} {\bibfnamefont {W.}~\bibnamefont
  {Hogervorst}}, \ and\ \bibinfo {author} {\bibfnamefont {W.}~\bibnamefont
  {Vassen}},\ }\href@noop {} {\bibfield  {journal} {\bibinfo  {journal}
  {Physical Review A}\ }\textbf {\bibinfo {volume} {73}},\ \bibinfo {pages}
  {032713} (\bibinfo {year} {2006})}\BibitemShut {NoStop}%
\bibitem [{\citenamefont {Busch}\ \emph {et~al.}(2006)\citenamefont {Busch},
  \citenamefont {Shaffer}, \citenamefont {Ahmed},\ and\ \citenamefont
  {Sukenik}}]{2006-ArRb}%
  \BibitemOpen
  \bibfield  {author} {\bibinfo {author} {\bibfnamefont {H.~C.}\ \bibnamefont
  {Busch}}, \bibinfo {author} {\bibfnamefont {M.~K.}\ \bibnamefont {Shaffer}},
  \bibinfo {author} {\bibfnamefont {E.~M.}\ \bibnamefont {Ahmed}}, \ and\
  \bibinfo {author} {\bibfnamefont {C.~I.}\ \bibnamefont {Sukenik}},\ }\href
  {\doibase 10.1103/PhysRevA.73.023406} {\bibfield  {journal} {\bibinfo
  {journal} {Phys. Rev. A}\ }\textbf {\bibinfo {volume} {73}},\ \bibinfo
  {pages} {023406} (\bibinfo {year} {2006})}\BibitemShut {NoStop}%
\bibitem [{\citenamefont {Glover}\ \emph {et~al.}(2011)\citenamefont {Glover},
  \citenamefont {Calvert}, \citenamefont {Laban},\ and\ \citenamefont
  {Sang}}]{2011-OpticalCollisions}%
  \BibitemOpen
  \bibfield  {author} {\bibinfo {author} {\bibfnamefont {R.}~\bibnamefont
  {Glover}}, \bibinfo {author} {\bibfnamefont {J.}~\bibnamefont {Calvert}},
  \bibinfo {author} {\bibfnamefont {D.}~\bibnamefont {Laban}}, \ and\ \bibinfo
  {author} {\bibfnamefont {R.}~\bibnamefont {Sang}},\ }\href@noop {} {\bibfield
   {journal} {\bibinfo  {journal} {Journal of Physics B: Atomic, Molecular and
  Optical Physics}\ }\textbf {\bibinfo {volume} {44}},\ \bibinfo {pages}
  {245202} (\bibinfo {year} {2011})}\BibitemShut {NoStop}%
\bibitem [{\citenamefont {Flores}\ \emph {et~al.}(2016)\citenamefont {Flores},
  \citenamefont {Vassen},\ and\ \citenamefont {Knoop}}]{2016-HeRb}%
  \BibitemOpen
  \bibfield  {author} {\bibinfo {author} {\bibfnamefont {A.~S.}\ \bibnamefont
  {Flores}}, \bibinfo {author} {\bibfnamefont {W.}~\bibnamefont {Vassen}}, \
  and\ \bibinfo {author} {\bibfnamefont {S.}~\bibnamefont {Knoop}},\ }\href
  {\doibase 10.1103/PhysRevA.94.050701} {\bibfield  {journal} {\bibinfo
  {journal} {Phys. Rev. A}\ }\textbf {\bibinfo {volume} {94}},\ \bibinfo
  {pages} {050701(R)} (\bibinfo {year} {2016})}\BibitemShut {NoStop}%
\bibitem [{\citenamefont {Mastwijk}\ \emph {et~al.}(1998)\citenamefont
  {Mastwijk}, \citenamefont {Thomsen}, \citenamefont {van~der Straten},\ and\
  \citenamefont {Niehaus}}]{1998-Mast}%
  \BibitemOpen
  \bibfield  {author} {\bibinfo {author} {\bibfnamefont {H.~C.}\ \bibnamefont
  {Mastwijk}}, \bibinfo {author} {\bibfnamefont {J.~W.}\ \bibnamefont
  {Thomsen}}, \bibinfo {author} {\bibfnamefont {P.}~\bibnamefont {van~der
  Straten}}, \ and\ \bibinfo {author} {\bibfnamefont {A.}~\bibnamefont
  {Niehaus}},\ }\href@noop {} {\bibfield  {journal} {\bibinfo  {journal}
  {Physical review letters}\ }\textbf {\bibinfo {volume} {80}},\ \bibinfo
  {pages} {5516} (\bibinfo {year} {1998})}\BibitemShut {NoStop}%
\bibitem [{\citenamefont {Deiglmayr}\ \emph {et~al.}(2008)\citenamefont
  {Deiglmayr}, \citenamefont {Grochola}, \citenamefont {Repp}, \citenamefont
  {M\"ortlbauer}, \citenamefont {Gl\"uck}, \citenamefont {Lange}, \citenamefont
  {Dulieu}, \citenamefont {Wester},\ and\ \citenamefont
  {Weidem\"uller}}]{2008-REMPI}%
  \BibitemOpen
  \bibfield  {author} {\bibinfo {author} {\bibfnamefont {J.}~\bibnamefont
  {Deiglmayr}}, \bibinfo {author} {\bibfnamefont {A.}~\bibnamefont {Grochola}},
  \bibinfo {author} {\bibfnamefont {M.}~\bibnamefont {Repp}}, \bibinfo {author}
  {\bibfnamefont {K.}~\bibnamefont {M\"ortlbauer}}, \bibinfo {author}
  {\bibfnamefont {C.}~\bibnamefont {Gl\"uck}}, \bibinfo {author} {\bibfnamefont
  {J.}~\bibnamefont {Lange}}, \bibinfo {author} {\bibfnamefont
  {O.}~\bibnamefont {Dulieu}}, \bibinfo {author} {\bibfnamefont
  {R.}~\bibnamefont {Wester}}, \ and\ \bibinfo {author} {\bibfnamefont
  {M.}~\bibnamefont {Weidem\"uller}},\ }\href {\doibase
  10.1103/PhysRevLett.101.133004} {\bibfield  {journal} {\bibinfo  {journal}
  {Phys. Rev. Lett.}\ }\textbf {\bibinfo {volume} {101}},\ \bibinfo {pages}
  {133004} (\bibinfo {year} {2008})}\BibitemShut {NoStop}%
\bibitem [{\citenamefont {Carini}\ \emph {et~al.}(2013)\citenamefont {Carini},
  \citenamefont {Pechkis}, \citenamefont {Rogers}, \citenamefont {Gould},
  \citenamefont {Kallush},\ and\ \citenamefont {Kosloff}}]{2013-REMPI}%
  \BibitemOpen
  \bibfield  {author} {\bibinfo {author} {\bibfnamefont {J.~L.}\ \bibnamefont
  {Carini}}, \bibinfo {author} {\bibfnamefont {J.~A.}\ \bibnamefont {Pechkis}},
  \bibinfo {author} {\bibfnamefont {C.~E.}\ \bibnamefont {Rogers}}, \bibinfo
  {author} {\bibfnamefont {P.~L.}\ \bibnamefont {Gould}}, \bibinfo {author}
  {\bibfnamefont {S.}~\bibnamefont {Kallush}}, \ and\ \bibinfo {author}
  {\bibfnamefont {R.}~\bibnamefont {Kosloff}},\ }\href {\doibase
  10.1103/PhysRevA.87.011401} {\bibfield  {journal} {\bibinfo  {journal} {Phys.
  Rev. A}\ }\textbf {\bibinfo {volume} {87}},\ \bibinfo {pages} {011401(R)}
  (\bibinfo {year} {2013})}\BibitemShut {NoStop}%
\bibitem [{\citenamefont {Bibelnik}\ \emph {et~al.}(2019)\citenamefont
  {Bibelnik}, \citenamefont {Gersten}, \citenamefont {Henson}, \citenamefont
  {Lavert-Ofir}, \citenamefont {Shagam}, \citenamefont {Skomorowski},
  \citenamefont {Koch},\ and\ \citenamefont {Narevicius}}]{2019-NeAr}%
  \BibitemOpen
  \bibfield  {author} {\bibinfo {author} {\bibfnamefont {N.}~\bibnamefont
  {Bibelnik}}, \bibinfo {author} {\bibfnamefont {S.}~\bibnamefont {Gersten}},
  \bibinfo {author} {\bibfnamefont {A.~B.}\ \bibnamefont {Henson}}, \bibinfo
  {author} {\bibfnamefont {E.}~\bibnamefont {Lavert-Ofir}}, \bibinfo {author}
  {\bibfnamefont {Y.}~\bibnamefont {Shagam}}, \bibinfo {author} {\bibfnamefont
  {W.}~\bibnamefont {Skomorowski}}, \bibinfo {author} {\bibfnamefont {C.~P.}\
  \bibnamefont {Koch}}, \ and\ \bibinfo {author} {\bibfnamefont
  {E.}~\bibnamefont {Narevicius}},\ }\href@noop {} {\bibfield  {journal}
  {\bibinfo  {journal} {Molecular Physics}\ ,\ \bibinfo {pages} {1}} (\bibinfo
  {year} {2019})}\BibitemShut {NoStop}%
\bibitem [{\citenamefont {Cop}\ and\ \citenamefont
  {Walser}(2018)}]{2018-Wasler}%
  \BibitemOpen
  \bibfield  {author} {\bibinfo {author} {\bibfnamefont {C.}~\bibnamefont
  {Cop}}\ and\ \bibinfo {author} {\bibfnamefont {R.}~\bibnamefont {Walser}},\
  }\href {\doibase 10.1103/PhysRevA.97.012704} {\bibfield  {journal} {\bibinfo
  {journal} {Phys. Rev. A}\ }\textbf {\bibinfo {volume} {97}},\ \bibinfo
  {pages} {012704} (\bibinfo {year} {2018})}\BibitemShut {NoStop}%
\bibitem [{\citenamefont {Cocks}\ \emph {et~al.}(2019)\citenamefont {Cocks},
  \citenamefont {Whittingham},\ and\ \citenamefont {Peach}}]{2019-UltracoldHe}%
  \BibitemOpen
  \bibfield  {author} {\bibinfo {author} {\bibfnamefont {D.~G.}\ \bibnamefont
  {Cocks}}, \bibinfo {author} {\bibfnamefont {I.~B.}\ \bibnamefont
  {Whittingham}}, \ and\ \bibinfo {author} {\bibfnamefont {G.}~\bibnamefont
  {Peach}},\ }\href {\doibase 10.1103/PhysRevA.99.062712} {\bibfield  {journal}
  {\bibinfo  {journal} {Phys. Rev. A}\ }\textbf {\bibinfo {volume} {99}},\
  \bibinfo {pages} {062712} (\bibinfo {year} {2019})}\BibitemShut {NoStop}%
\bibitem [{\citenamefont {Siska}(1993)}]{1993-Siska}%
  \BibitemOpen
  \bibfield  {author} {\bibinfo {author} {\bibfnamefont {P.~E.}\ \bibnamefont
  {Siska}},\ }\href {\doibase 10.1103/RevModPhys.65.337} {\bibfield  {journal}
  {\bibinfo  {journal} {Rev. Mod. Phys.}\ }\textbf {\bibinfo {volume} {65}},\
  \bibinfo {pages} {337} (\bibinfo {year} {1993})}\BibitemShut {NoStop}%
\bibitem [{\citenamefont {Dörner}\ \emph {et~al.}(2000)\citenamefont
  {Dörner}, \citenamefont {Mergel}, \citenamefont {Jagutzki}, \citenamefont
  {Spielberger}, \citenamefont {Ullrich}, \citenamefont {Moshammer},\ and\
  \citenamefont {Schmidt-Böcking}}]{2000-COLTRIMS_Review}%
  \BibitemOpen
  \bibfield  {author} {\bibinfo {author} {\bibfnamefont {R.}~\bibnamefont
  {Dörner}}, \bibinfo {author} {\bibfnamefont {V.}~\bibnamefont {Mergel}},
  \bibinfo {author} {\bibfnamefont {O.}~\bibnamefont {Jagutzki}}, \bibinfo
  {author} {\bibfnamefont {L.}~\bibnamefont {Spielberger}}, \bibinfo {author}
  {\bibfnamefont {J.}~\bibnamefont {Ullrich}}, \bibinfo {author} {\bibfnamefont
  {R.}~\bibnamefont {Moshammer}}, \ and\ \bibinfo {author} {\bibfnamefont
  {H.}~\bibnamefont {Schmidt-Böcking}},\ }\href {\doibase
  https://doi.org/10.1016/S0370-1573(99)00109-X} {\bibfield  {journal}
  {\bibinfo  {journal} {Physics Reports}\ }\textbf {\bibinfo {volume} {330}},\
  \bibinfo {pages} {95 } (\bibinfo {year} {2000})}\BibitemShut {NoStop}%
\bibitem [{\citenamefont {Eppink}\ and\ \citenamefont
  {Parker}(1997)}]{1997-EP}%
  \BibitemOpen
  \bibfield  {author} {\bibinfo {author} {\bibfnamefont {A.~T. J.~B.}\
  \bibnamefont {Eppink}}\ and\ \bibinfo {author} {\bibfnamefont {D.~H.}\
  \bibnamefont {Parker}},\ }\href {\doibase 10.1063/1.1148310} {\bibfield
  {journal} {\bibinfo  {journal} {Review of Scientific Instruments}\ }\textbf
  {\bibinfo {volume} {68}},\ \bibinfo {pages} {3477} (\bibinfo {year}
  {1997})},\ \Eprint {http://arxiv.org/abs/https://doi.org/10.1063/1.1148310}
  {https://doi.org/10.1063/1.1148310} \BibitemShut {NoStop}%
\bibitem [{\citenamefont {Chandler}\ \emph {et~al.}(2017)\citenamefont
  {Chandler}, \citenamefont {Houston},\ and\ \citenamefont
  {Parker}}]{2017-Parker}%
  \BibitemOpen
  \bibfield  {author} {\bibinfo {author} {\bibfnamefont {D.~W.}\ \bibnamefont
  {Chandler}}, \bibinfo {author} {\bibfnamefont {P.~L.}\ \bibnamefont
  {Houston}}, \ and\ \bibinfo {author} {\bibfnamefont {D.~H.}\ \bibnamefont
  {Parker}},\ }\href {\doibase 10.1063/1.4983623} {\bibfield  {journal}
  {\bibinfo  {journal} {The Journal of Chemical Physics}\ }\textbf {\bibinfo
  {volume} {147}},\ \bibinfo {pages} {013601} (\bibinfo {year} {2017})},\
  \Eprint {http://arxiv.org/abs/https://doi.org/10.1063/1.4983623}
  {https://doi.org/10.1063/1.4983623} \BibitemShut {NoStop}%
\bibitem [{\citenamefont {Arango}\ \emph {et~al.}(2006)\citenamefont {Arango},
  \citenamefont {Shapiro},\ and\ \citenamefont {Brumer}}]{2006-Coherent}%
  \BibitemOpen
  \bibfield  {author} {\bibinfo {author} {\bibfnamefont {C.~A.}\ \bibnamefont
  {Arango}}, \bibinfo {author} {\bibfnamefont {M.}~\bibnamefont {Shapiro}}, \
  and\ \bibinfo {author} {\bibfnamefont {P.}~\bibnamefont {Brumer}},\
  }\href@noop {} {\bibfield  {journal} {\bibinfo  {journal} {Physical review
  letters}\ }\textbf {\bibinfo {volume} {97}},\ \bibinfo {pages} {193202}
  (\bibinfo {year} {2006})}\BibitemShut {NoStop}%
\bibitem [{\citenamefont {Behr}\ and\ \citenamefont
  {Gwinner}(2009)}]{2009-BehrTraps}%
  \BibitemOpen
  \bibfield  {author} {\bibinfo {author} {\bibfnamefont {J.}~\bibnamefont
  {Behr}}\ and\ \bibinfo {author} {\bibfnamefont {G.}~\bibnamefont {Gwinner}},\
  }\href@noop {} {\bibfield  {journal} {\bibinfo  {journal} {Journal of Physics
  G: Nuclear and Particle Physics}\ }\textbf {\bibinfo {volume} {36}},\
  \bibinfo {pages} {033101} (\bibinfo {year} {2009})}\BibitemShut {NoStop}%
\bibitem [{\citenamefont {Yee}\ \emph {et~al.}(2013)\citenamefont {Yee},
  \citenamefont {Scielzo}, \citenamefont {Bertone}, \citenamefont {Buchinger},
  \citenamefont {Caldwell}, \citenamefont {Clark}, \citenamefont {Deibel},
  \citenamefont {Fallis}, \citenamefont {Greene}, \citenamefont {Gulick},
  \citenamefont {Lascar}, \citenamefont {Levand}, \citenamefont {Li},
  \citenamefont {Norman}, \citenamefont {Pedretti}, \citenamefont {Savard},
  \citenamefont {Segel}, \citenamefont {Sharma}, \citenamefont {Sternberg},
  \citenamefont {VanSchelt},\ and\ \citenamefont {Zabransky}}]{2013-Delayed}%
  \BibitemOpen
  \bibfield  {author} {\bibinfo {author} {\bibfnamefont {R.~M.}\ \bibnamefont
  {Yee}}, \bibinfo {author} {\bibfnamefont {N.~D.}\ \bibnamefont {Scielzo}},
  \bibinfo {author} {\bibfnamefont {P.~F.}\ \bibnamefont {Bertone}}, \bibinfo
  {author} {\bibfnamefont {F.}~\bibnamefont {Buchinger}}, \bibinfo {author}
  {\bibfnamefont {S.}~\bibnamefont {Caldwell}}, \bibinfo {author}
  {\bibfnamefont {J.~A.}\ \bibnamefont {Clark}}, \bibinfo {author}
  {\bibfnamefont {C.~M.}\ \bibnamefont {Deibel}}, \bibinfo {author}
  {\bibfnamefont {J.}~\bibnamefont {Fallis}}, \bibinfo {author} {\bibfnamefont
  {J.~P.}\ \bibnamefont {Greene}}, \bibinfo {author} {\bibfnamefont
  {S.}~\bibnamefont {Gulick}}, \bibinfo {author} {\bibfnamefont
  {D.}~\bibnamefont {Lascar}}, \bibinfo {author} {\bibfnamefont {A.~F.}\
  \bibnamefont {Levand}}, \bibinfo {author} {\bibfnamefont {G.}~\bibnamefont
  {Li}}, \bibinfo {author} {\bibfnamefont {E.~B.}\ \bibnamefont {Norman}},
  \bibinfo {author} {\bibfnamefont {M.}~\bibnamefont {Pedretti}}, \bibinfo
  {author} {\bibfnamefont {G.}~\bibnamefont {Savard}}, \bibinfo {author}
  {\bibfnamefont {R.~E.}\ \bibnamefont {Segel}}, \bibinfo {author}
  {\bibfnamefont {K.~S.}\ \bibnamefont {Sharma}}, \bibinfo {author}
  {\bibfnamefont {M.~G.}\ \bibnamefont {Sternberg}}, \bibinfo {author}
  {\bibfnamefont {J.}~\bibnamefont {VanSchelt}}, \ and\ \bibinfo {author}
  {\bibfnamefont {B.~J.}\ \bibnamefont {Zabransky}},\ }\href@noop {} {\bibfield
   {journal} {\bibinfo  {journal} {Physical review letters}\ }\textbf {\bibinfo
  {volume} {110}},\ \bibinfo {pages} {092501} (\bibinfo {year}
  {2013})}\BibitemShut {NoStop}%
\bibitem [{\citenamefont {Br{\'e}dy}\ \emph {et~al.}(2003)\citenamefont
  {Br{\'e}dy}, \citenamefont {Nguyen}, \citenamefont {Camp}, \citenamefont
  {Fl{\'e}chard},\ and\ \citenamefont {DePaola}}]{2003-MOTRIMS}%
  \BibitemOpen
  \bibfield  {author} {\bibinfo {author} {\bibfnamefont {R.}~\bibnamefont
  {Br{\'e}dy}}, \bibinfo {author} {\bibfnamefont {H.}~\bibnamefont {Nguyen}},
  \bibinfo {author} {\bibfnamefont {H.}~\bibnamefont {Camp}}, \bibinfo {author}
  {\bibfnamefont {X.}~\bibnamefont {Fl{\'e}chard}}, \ and\ \bibinfo {author}
  {\bibfnamefont {B.}~\bibnamefont {DePaola}},\ }\href@noop {} {\bibfield
  {journal} {\bibinfo  {journal} {Nuclear Instruments and Methods in Physics
  Research Section B: Beam Interactions with Materials and Atoms}\ }\textbf
  {\bibinfo {volume} {205}},\ \bibinfo {pages} {191} (\bibinfo {year}
  {2003})}\BibitemShut {NoStop}%
\bibitem [{\citenamefont {Weger}\ \emph {et~al.}(2013)\citenamefont {Weger},
  \citenamefont {Maurer}, \citenamefont {Ludwig}, \citenamefont {Gallmann},\
  and\ \citenamefont {Keller}}]{2013-COLTRIM_VMI_Comp}%
  \BibitemOpen
  \bibfield  {author} {\bibinfo {author} {\bibfnamefont {M.}~\bibnamefont
  {Weger}}, \bibinfo {author} {\bibfnamefont {J.}~\bibnamefont {Maurer}},
  \bibinfo {author} {\bibfnamefont {A.}~\bibnamefont {Ludwig}}, \bibinfo
  {author} {\bibfnamefont {L.}~\bibnamefont {Gallmann}}, \ and\ \bibinfo
  {author} {\bibfnamefont {U.}~\bibnamefont {Keller}},\ }\href@noop {}
  {\bibfield  {journal} {\bibinfo  {journal} {Optics express}\ }\textbf
  {\bibinfo {volume} {21}},\ \bibinfo {pages} {21981} (\bibinfo {year}
  {2013})}\BibitemShut {NoStop}%
\bibitem [{\citenamefont {Kling}\ \emph {et~al.}(2014)\citenamefont {Kling},
  \citenamefont {Paul}, \citenamefont {Gura}, \citenamefont {Laurent},
  \citenamefont {De}, \citenamefont {Li}, \citenamefont {Wang}, \citenamefont
  {Ahn}, \citenamefont {Kim}, \citenamefont {Kim} \emph
  {et~al.}}]{2014-Kling_Thick}%
  \BibitemOpen
  \bibfield  {author} {\bibinfo {author} {\bibfnamefont {N.}~\bibnamefont
  {Kling}}, \bibinfo {author} {\bibfnamefont {D.}~\bibnamefont {Paul}},
  \bibinfo {author} {\bibfnamefont {A.}~\bibnamefont {Gura}}, \bibinfo {author}
  {\bibfnamefont {G.}~\bibnamefont {Laurent}}, \bibinfo {author} {\bibfnamefont
  {S.}~\bibnamefont {De}}, \bibinfo {author} {\bibfnamefont {H.}~\bibnamefont
  {Li}}, \bibinfo {author} {\bibfnamefont {Z.}~\bibnamefont {Wang}}, \bibinfo
  {author} {\bibfnamefont {B.}~\bibnamefont {Ahn}}, \bibinfo {author}
  {\bibfnamefont {C.}~\bibnamefont {Kim}}, \bibinfo {author} {\bibfnamefont
  {T.}~\bibnamefont {Kim}},  \emph {et~al.},\ }\href@noop {} {\bibfield
  {journal} {\bibinfo  {journal} {Journal of Instrumentation}\ }\textbf
  {\bibinfo {volume} {9}},\ \bibinfo {pages} {P05005} (\bibinfo {year}
  {2014})}\BibitemShut {NoStop}%
\bibitem [{\citenamefont {DePaola}\ \emph {et~al.}(2008)\citenamefont
  {DePaola}, \citenamefont {Morgenstern},\ and\ \citenamefont
  {Andersen}}]{2008-MOTRIMS}%
  \BibitemOpen
  \bibfield  {author} {\bibinfo {author} {\bibfnamefont {B.}~\bibnamefont
  {DePaola}}, \bibinfo {author} {\bibfnamefont {R.}~\bibnamefont
  {Morgenstern}}, \ and\ \bibinfo {author} {\bibfnamefont {N.}~\bibnamefont
  {Andersen}},\ }\href@noop {} {\bibfield  {journal} {\bibinfo  {journal}
  {Advances In Atomic, Molecular, and Optical Physics}\ }\textbf {\bibinfo
  {volume} {55}},\ \bibinfo {pages} {139} (\bibinfo {year} {2008})}\BibitemShut
  {NoStop}%
\bibitem [{\citenamefont {Blieck}\ \emph {et~al.}(2009)\citenamefont {Blieck},
  \citenamefont {Fl{\'e}chard}, \citenamefont {Cassimi}, \citenamefont
  {Gilles}, \citenamefont {Girard},\ and\ \citenamefont
  {Hennecart}}]{2009-MOTIMRS}%
  \BibitemOpen
  \bibfield  {author} {\bibinfo {author} {\bibfnamefont {J.}~\bibnamefont
  {Blieck}}, \bibinfo {author} {\bibfnamefont {X.}~\bibnamefont
  {Fl{\'e}chard}}, \bibinfo {author} {\bibfnamefont {A.}~\bibnamefont
  {Cassimi}}, \bibinfo {author} {\bibfnamefont {H.}~\bibnamefont {Gilles}},
  \bibinfo {author} {\bibfnamefont {S.}~\bibnamefont {Girard}}, \ and\ \bibinfo
  {author} {\bibfnamefont {D.}~\bibnamefont {Hennecart}},\ }in\ \href@noop {}
  {\emph {\bibinfo {booktitle} {Journal of Physics: Conference Series}}},\
  Vol.\ \bibinfo {volume} {163}\ (\bibinfo {organization} {IOP Publishing},\
  \bibinfo {year} {2009})\ p.\ \bibinfo {pages} {012070}\BibitemShut {NoStop}%
\bibitem [{\citenamefont {Ohayon}\ and\ \citenamefont
  {Ron}(2015)}]{2015-Zeeman}%
  \BibitemOpen
  \bibfield  {author} {\bibinfo {author} {\bibfnamefont {B.}~\bibnamefont
  {Ohayon}}\ and\ \bibinfo {author} {\bibfnamefont {G.}~\bibnamefont {Ron}},\
  }\href@noop {} {\bibfield  {journal} {\bibinfo  {journal} {Review of
  Scientific Instruments}\ }\textbf {\bibinfo {volume} {86}},\ \bibinfo {pages}
  {103110} (\bibinfo {year} {2015})}\BibitemShut {NoStop}%
\bibitem [{\citenamefont {Mardor}\ \emph {et~al.}(2018)\citenamefont {Mardor},
  \citenamefont {Aviv}, \citenamefont {Avrigeanu}, \citenamefont {Berkovits},
  \citenamefont {Dahan}, \citenamefont {Dickel}, \citenamefont {Eliyahu},
  \citenamefont {Gai}, \citenamefont {Gavish-Segev}, \citenamefont {Halfon}
  \emph {et~al.}}]{2018-Mardor}%
  \BibitemOpen
  \bibfield  {author} {\bibinfo {author} {\bibfnamefont {I.}~\bibnamefont
  {Mardor}}, \bibinfo {author} {\bibfnamefont {O.}~\bibnamefont {Aviv}},
  \bibinfo {author} {\bibfnamefont {M.}~\bibnamefont {Avrigeanu}}, \bibinfo
  {author} {\bibfnamefont {D.}~\bibnamefont {Berkovits}}, \bibinfo {author}
  {\bibfnamefont {A.}~\bibnamefont {Dahan}}, \bibinfo {author} {\bibfnamefont
  {T.}~\bibnamefont {Dickel}}, \bibinfo {author} {\bibfnamefont
  {I.}~\bibnamefont {Eliyahu}}, \bibinfo {author} {\bibfnamefont
  {M.}~\bibnamefont {Gai}}, \bibinfo {author} {\bibfnamefont {I.}~\bibnamefont
  {Gavish-Segev}}, \bibinfo {author} {\bibfnamefont {S.}~\bibnamefont
  {Halfon}},  \emph {et~al.},\ }\href@noop {} {\bibfield  {journal} {\bibinfo
  {journal} {The European Physical Journal A}\ }\textbf {\bibinfo {volume}
  {54}},\ \bibinfo {pages} {91} (\bibinfo {year} {2018})}\BibitemShut {NoStop}%
\bibitem [{\citenamefont {Sanders}\ and\ \citenamefont
  {Muschlitz}(1977)}]{1976-Cehniionization}%
  \BibitemOpen
  \bibfield  {author} {\bibinfo {author} {\bibfnamefont {R.}~\bibnamefont
  {Sanders}}\ and\ \bibinfo {author} {\bibfnamefont {E.}~\bibnamefont
  {Muschlitz}},\ }\href {\doibase https://doi.org/10.1016/0020-7381(77)80092-2}
  {\bibfield  {journal} {\bibinfo  {journal} {International Journal of Mass
  Spectrometry and Ion Physics}\ }\textbf {\bibinfo {volume} {23}},\ \bibinfo
  {pages} {99 } (\bibinfo {year} {1977})}\BibitemShut {NoStop}%
\bibitem [{\citenamefont {Falcinelli}\ \emph {et~al.}(2016)\citenamefont
  {Falcinelli}, \citenamefont {Rosi}, \citenamefont {Pirani}, \citenamefont
  {Stranges},\ and\ \citenamefont {Vecchiocattivi}}]{2016-NeH2O}%
  \BibitemOpen
  \bibfield  {author} {\bibinfo {author} {\bibfnamefont {S.}~\bibnamefont
  {Falcinelli}}, \bibinfo {author} {\bibfnamefont {M.}~\bibnamefont {Rosi}},
  \bibinfo {author} {\bibfnamefont {F.}~\bibnamefont {Pirani}}, \bibinfo
  {author} {\bibfnamefont {D.}~\bibnamefont {Stranges}}, \ and\ \bibinfo
  {author} {\bibfnamefont {F.}~\bibnamefont {Vecchiocattivi}},\ }\href@noop {}
  {\bibfield  {journal} {\bibinfo  {journal} {The Journal of Physical Chemistry
  A}\ }\textbf {\bibinfo {volume} {120}},\ \bibinfo {pages} {5169} (\bibinfo
  {year} {2016})}\BibitemShut {NoStop}%
\bibitem [{\citenamefont {Balucani}\ \emph {et~al.}(2012)\citenamefont
  {Balucani}, \citenamefont {Bartocci}, \citenamefont {Brunetti}, \citenamefont
  {Candori}, \citenamefont {Falcinelli}, \citenamefont {Palazzetti},
  \citenamefont {Pirani},\ and\ \citenamefont {Vecchiocattivi}}]{2012-H2O}%
  \BibitemOpen
  \bibfield  {author} {\bibinfo {author} {\bibfnamefont {N.}~\bibnamefont
  {Balucani}}, \bibinfo {author} {\bibfnamefont {A.}~\bibnamefont {Bartocci}},
  \bibinfo {author} {\bibfnamefont {B.}~\bibnamefont {Brunetti}}, \bibinfo
  {author} {\bibfnamefont {P.}~\bibnamefont {Candori}}, \bibinfo {author}
  {\bibfnamefont {S.}~\bibnamefont {Falcinelli}}, \bibinfo {author}
  {\bibfnamefont {F.}~\bibnamefont {Palazzetti}}, \bibinfo {author}
  {\bibfnamefont {F.}~\bibnamefont {Pirani}}, \ and\ \bibinfo {author}
  {\bibfnamefont {F.}~\bibnamefont {Vecchiocattivi}},\ }\href@noop {}
  {\bibfield  {journal} {\bibinfo  {journal} {Chemical Physics Letters}\
  }\textbf {\bibinfo {volume} {546}},\ \bibinfo {pages} {34} (\bibinfo {year}
  {2012})}\BibitemShut {NoStop}%
\bibitem [{\citenamefont {Brunetti}\ \emph {et~al.}(2013)\citenamefont
  {Brunetti}, \citenamefont {Candori}, \citenamefont {Falcinelli},
  \citenamefont {Pirani},\ and\ \citenamefont {Vecchiocattivi}}]{2013-Stero}%
  \BibitemOpen
  \bibfield  {author} {\bibinfo {author} {\bibfnamefont {B.~G.}\ \bibnamefont
  {Brunetti}}, \bibinfo {author} {\bibfnamefont {P.}~\bibnamefont {Candori}},
  \bibinfo {author} {\bibfnamefont {S.}~\bibnamefont {Falcinelli}}, \bibinfo
  {author} {\bibfnamefont {F.}~\bibnamefont {Pirani}}, \ and\ \bibinfo {author}
  {\bibfnamefont {F.}~\bibnamefont {Vecchiocattivi}},\ }\href@noop {}
  {\bibfield  {journal} {\bibinfo  {journal} {The Journal of chemical physics}\
  }\textbf {\bibinfo {volume} {139}},\ \bibinfo {pages} {164305} (\bibinfo
  {year} {2013})}\BibitemShut {NoStop}%
\bibitem [{Sup()}]{Sup}%
  \BibitemOpen
  \href@noop {} {}\bibinfo {note} {See Supplemental Material at... for a
  discussion on the probability of spontaneous emissions, the extraction of
  branching ratios for various laser intensities, and details regarding the
  inversion process.}\BibitemShut {Stop}%
\bibitem [{\citenamefont {Oberheide}\ \emph
  {et~al.}(1997{\natexlab{a}})\citenamefont {Oberheide}, \citenamefont
  {Wilhelms},\ and\ \citenamefont {Zimmer}}]{1997-MCPsup}%
  \BibitemOpen
  \bibfield  {author} {\bibinfo {author} {\bibfnamefont {J.}~\bibnamefont
  {Oberheide}}, \bibinfo {author} {\bibfnamefont {P.}~\bibnamefont {Wilhelms}},
  \ and\ \bibinfo {author} {\bibfnamefont {M.}~\bibnamefont {Zimmer}},\
  }\href@noop {} {\bibfield  {journal} {\bibinfo  {journal} {Measurement
  Science and Technology}\ }\textbf {\bibinfo {volume} {8}},\ \bibinfo {pages}
  {351} (\bibinfo {year} {1997}{\natexlab{a}})}\BibitemShut {NoStop}%
\bibitem [{\citenamefont {Ohayon}\ \emph
  {et~al.}(2019{\natexlab{a}})\citenamefont {Ohayon}, \citenamefont
  {Rahangdale}, \citenamefont {Geddes}, \citenamefont {Berengut},\ and\
  \citenamefont {Ron}}]{2019-ISsup}%
  \BibitemOpen
  \bibfield  {author} {\bibinfo {author} {\bibfnamefont {B.}~\bibnamefont
  {Ohayon}}, \bibinfo {author} {\bibfnamefont {H.}~\bibnamefont {Rahangdale}},
  \bibinfo {author} {\bibfnamefont {A.~J.}\ \bibnamefont {Geddes}}, \bibinfo
  {author} {\bibfnamefont {J.~C.}\ \bibnamefont {Berengut}}, \ and\ \bibinfo
  {author} {\bibfnamefont {G.}~\bibnamefont {Ron}},\ }\href {\doibase
  10.1103/PhysRevA.99.042503} {\bibfield  {journal} {\bibinfo  {journal} {Phys.
  Rev. A}\ }\textbf {\bibinfo {volume} {99}},\ \bibinfo {pages} {042503}
  (\bibinfo {year} {2019}{\natexlab{a}})}\BibitemShut {NoStop}%
\bibitem [{\citenamefont {Ohayon}\ \emph
  {et~al.}(2015{\natexlab{a}})\citenamefont {Ohayon}, \citenamefont
  {W{\aa}hlin},\ and\ \citenamefont {Ron}}]{2015-RFsup}%
  \BibitemOpen
  \bibfield  {author} {\bibinfo {author} {\bibfnamefont {B.}~\bibnamefont
  {Ohayon}}, \bibinfo {author} {\bibfnamefont {E.}~\bibnamefont {W{\aa}hlin}},
  \ and\ \bibinfo {author} {\bibfnamefont {G.}~\bibnamefont {Ron}},\
  }\href@noop {} {\bibfield  {journal} {\bibinfo  {journal} {Journal of
  Instrumentation}\ }\textbf {\bibinfo {volume} {10}},\ \bibinfo {pages}
  {P03009} (\bibinfo {year} {2015}{\natexlab{a}})}\BibitemShut {NoStop}%
\bibitem [{\citenamefont {Van~Drunen}(2008)}]{2008-Drunen}%
  \BibitemOpen
  \bibfield  {author} {\bibinfo {author} {\bibfnamefont {W.~J.}\ \bibnamefont
  {Van~Drunen}},\ }\emph {\bibinfo {title} {Collisional interaction between
  metastable neon atoms}},\ \href@noop {} {Ph.D. thesis},\ \bibinfo  {school}
  {Technische Universit{\"a}t Darmstadt} (\bibinfo {year} {2008})\BibitemShut
  {NoStop}%
\bibitem [{\citenamefont {Tol}\ \emph {et~al.}(1999)\citenamefont {Tol},
  \citenamefont {Herschbach}, \citenamefont {Hessels}, \citenamefont
  {Hogervorst},\ and\ \citenamefont {Vassen}}]{1999-Tol}%
  \BibitemOpen
  \bibfield  {author} {\bibinfo {author} {\bibfnamefont {P.~J.~J.}\
  \bibnamefont {Tol}}, \bibinfo {author} {\bibfnamefont {N.}~\bibnamefont
  {Herschbach}}, \bibinfo {author} {\bibfnamefont {E.~A.}\ \bibnamefont
  {Hessels}}, \bibinfo {author} {\bibfnamefont {W.}~\bibnamefont {Hogervorst}},
  \ and\ \bibinfo {author} {\bibfnamefont {W.}~\bibnamefont {Vassen}},\
  }\href@noop {} {\bibfield  {journal} {\bibinfo  {journal} {Physical Review
  A}\ }\textbf {\bibinfo {volume} {60}},\ \bibinfo {pages} {R761} (\bibinfo
  {year} {1999})}\BibitemShut {NoStop}%
\bibitem [{\citenamefont {Whitaker}(2003{\natexlab{a}})}]{2003-imagingSup}%
  \BibitemOpen
  \bibfield  {author} {\bibinfo {author} {\bibfnamefont {B.~J.}\ \bibnamefont
  {Whitaker}},\ }\href@noop {} {\emph {\bibinfo {title} {Imaging in molecular
  dynamics: technology and applications}}}\ (\bibinfo  {publisher} {Cambridge
  university press},\ \bibinfo {year} {2003})\BibitemShut {NoStop}%
\bibitem [{\citenamefont {Pieksma}\ \emph {et~al.}(2002)\citenamefont
  {Pieksma}, \citenamefont {Cizek}, \citenamefont {Thomsen}, \citenamefont
  {van~der Straten},\ and\ \citenamefont {Niehaus}}]{2002-RIMSHe}%
  \BibitemOpen
  \bibfield  {author} {\bibinfo {author} {\bibfnamefont {M.}~\bibnamefont
  {Pieksma}}, \bibinfo {author} {\bibfnamefont {M.}~\bibnamefont {Cizek}},
  \bibinfo {author} {\bibfnamefont {J.~W.}\ \bibnamefont {Thomsen}}, \bibinfo
  {author} {\bibfnamefont {P.}~\bibnamefont {van~der Straten}}, \ and\ \bibinfo
  {author} {\bibfnamefont {A.}~\bibnamefont {Niehaus}},\ }\href@noop {}
  {\bibfield  {journal} {\bibinfo  {journal} {Physical Review A}\ }\textbf
  {\bibinfo {volume} {66}},\ \bibinfo {pages} {022703} (\bibinfo {year}
  {2002})}\BibitemShut {NoStop}%
\bibitem [{\citenamefont {Miller}(1970)}]{1970-Miller}%
  \BibitemOpen
  \bibfield  {author} {\bibinfo {author} {\bibfnamefont {W.~H.}\ \bibnamefont
  {Miller}},\ }\href@noop {} {\bibfield  {journal} {\bibinfo  {journal} {The
  Journal of Chemical Physics}\ }\textbf {\bibinfo {volume} {52}},\ \bibinfo
  {pages} {3563} (\bibinfo {year} {1970})}\BibitemShut {NoStop}%
\bibitem [{\citenamefont {Hotop}(1974)}]{1974-Hotop}%
  \BibitemOpen
  \bibfield  {author} {\bibinfo {author} {\bibfnamefont {H.}~\bibnamefont
  {Hotop}},\ }\href@noop {} {\bibfield  {journal} {\bibinfo  {journal}
  {Radiation research}\ }\textbf {\bibinfo {volume} {59}},\ \bibinfo {pages}
  {379} (\bibinfo {year} {1974})}\BibitemShut {NoStop}%
\bibitem [{\citenamefont {Doery}\ \emph
  {et~al.}(1998{\natexlab{a}})\citenamefont {Doery}, \citenamefont
  {Vredenbregt}, \citenamefont {Op~de Beek}, \citenamefont {Beijerinck},\ and\
  \citenamefont {Verhaar}}]{1998-Doery}%
  \BibitemOpen
  \bibfield  {author} {\bibinfo {author} {\bibfnamefont {M.~R.}\ \bibnamefont
  {Doery}}, \bibinfo {author} {\bibfnamefont {E.~J.~D.}\ \bibnamefont
  {Vredenbregt}}, \bibinfo {author} {\bibfnamefont {S.~S.}\ \bibnamefont {Op~de
  Beek}}, \bibinfo {author} {\bibfnamefont {H.~C.~W.}\ \bibnamefont
  {Beijerinck}}, \ and\ \bibinfo {author} {\bibfnamefont {B.~J.}\ \bibnamefont
  {Verhaar}},\ }\href@noop {} {\bibfield  {journal} {\bibinfo  {journal}
  {Physical Review A}\ }\textbf {\bibinfo {volume} {58}},\ \bibinfo {pages}
  {3673} (\bibinfo {year} {1998}{\natexlab{a}})}\BibitemShut {NoStop}%
\bibitem [{\citenamefont {Kotochigova}\ \emph {et~al.}(2000)\citenamefont
  {Kotochigova}, \citenamefont {Tiesinga},\ and\ \citenamefont
  {Tupitsyn}}]{2000-Tupitsyn}%
  \BibitemOpen
  \bibfield  {author} {\bibinfo {author} {\bibfnamefont {S.}~\bibnamefont
  {Kotochigova}}, \bibinfo {author} {\bibfnamefont {E.}~\bibnamefont
  {Tiesinga}}, \ and\ \bibinfo {author} {\bibfnamefont {I.}~\bibnamefont
  {Tupitsyn}},\ }\href@noop {} {\bibfield  {journal} {\bibinfo  {journal}
  {Physical Review A}\ }\textbf {\bibinfo {volume} {61}},\ \bibinfo {pages}
  {042712} (\bibinfo {year} {2000})}\BibitemShut {NoStop}%
\bibitem [{\citenamefont {Trevor}\ \emph {et~al.}(1984)\citenamefont {Trevor},
  \citenamefont {Pollard}, \citenamefont {Brewer}, \citenamefont {Southworth},
  \citenamefont {Truesdale}, \citenamefont {Shirley},\ and\ \citenamefont
  {Lee}}]{1984-Photoionzation}%
  \BibitemOpen
  \bibfield  {author} {\bibinfo {author} {\bibfnamefont {D.}~\bibnamefont
  {Trevor}}, \bibinfo {author} {\bibfnamefont {J.}~\bibnamefont {Pollard}},
  \bibinfo {author} {\bibfnamefont {W.}~\bibnamefont {Brewer}}, \bibinfo
  {author} {\bibfnamefont {S.}~\bibnamefont {Southworth}}, \bibinfo {author}
  {\bibfnamefont {C.}~\bibnamefont {Truesdale}}, \bibinfo {author}
  {\bibfnamefont {D.}~\bibnamefont {Shirley}}, \ and\ \bibinfo {author}
  {\bibfnamefont {Y.}~\bibnamefont {Lee}},\ }\href@noop {} {\bibfield
  {journal} {\bibinfo  {journal} {The Journal of chemical physics}\ }\textbf
  {\bibinfo {volume} {80}},\ \bibinfo {pages} {6083} (\bibinfo {year}
  {1984})}\BibitemShut {NoStop}%
\bibitem [{\citenamefont {Carrington}\ \emph {et~al.}(2002)\citenamefont
  {Carrington}, \citenamefont {Gammie}, \citenamefont {Page}, \citenamefont
  {Shaw},\ and\ \citenamefont {Hutson}}]{2002-Microwave}%
  \BibitemOpen
  \bibfield  {author} {\bibinfo {author} {\bibfnamefont {A.}~\bibnamefont
  {Carrington}}, \bibinfo {author} {\bibfnamefont {D.~I.}\ \bibnamefont
  {Gammie}}, \bibinfo {author} {\bibfnamefont {J.~C.}\ \bibnamefont {Page}},
  \bibinfo {author} {\bibfnamefont {A.~M.}\ \bibnamefont {Shaw}}, \ and\
  \bibinfo {author} {\bibfnamefont {J.~M.}\ \bibnamefont {Hutson}},\
  }\href@noop {} {\bibfield  {journal} {\bibinfo  {journal} {The Journal of
  chemical physics}\ }\textbf {\bibinfo {volume} {116}},\ \bibinfo {pages}
  {3662} (\bibinfo {year} {2002})}\BibitemShut {NoStop}%
\bibitem [{\citenamefont {Doery}\ \emph
  {et~al.}(1998{\natexlab{b}})\citenamefont {Doery}, \citenamefont
  {Vredenbregt}, \citenamefont {Tempelaars}, \citenamefont {Beijerinck},\ and\
  \citenamefont {Verhaar}}]{1988-DoerySup}%
  \BibitemOpen
  \bibfield  {author} {\bibinfo {author} {\bibfnamefont {M.~R.}\ \bibnamefont
  {Doery}}, \bibinfo {author} {\bibfnamefont {E.~J.~D.}\ \bibnamefont
  {Vredenbregt}}, \bibinfo {author} {\bibfnamefont {J.~G.~C.}\ \bibnamefont
  {Tempelaars}}, \bibinfo {author} {\bibfnamefont {H.~C.~W.}\ \bibnamefont
  {Beijerinck}}, \ and\ \bibinfo {author} {\bibfnamefont {B.~J.}\ \bibnamefont
  {Verhaar}},\ }\href {\doibase 10.1103/PhysRevA.57.3603} {\bibfield  {journal}
  {\bibinfo  {journal} {Phys. Rev. A}\ }\textbf {\bibinfo {volume} {57}},\
  \bibinfo {pages} {3603} (\bibinfo {year} {1998}{\natexlab{b}})}\BibitemShut
  {NoStop}%
\bibitem [{\citenamefont {Gallagher}\ and\ \citenamefont
  {Pritchard}(1989{\natexlab{a}})}]{1989-GPsup}%
  \BibitemOpen
  \bibfield  {author} {\bibinfo {author} {\bibfnamefont {A.}~\bibnamefont
  {Gallagher}}\ and\ \bibinfo {author} {\bibfnamefont {D.~E.}\ \bibnamefont
  {Pritchard}},\ }\href@noop {} {\bibfield  {journal} {\bibinfo  {journal}
  {Physical review letters}\ }\textbf {\bibinfo {volume} {63}},\ \bibinfo
  {pages} {957} (\bibinfo {year} {1989}{\natexlab{a}})}\BibitemShut {NoStop}%
\bibitem [{\citenamefont {Carini}\ \emph {et~al.}(2015)\citenamefont {Carini},
  \citenamefont {Kallush}, \citenamefont {Kosloff},\ and\ \citenamefont
  {Gould}}]{2015-Nanosecond}%
  \BibitemOpen
  \bibfield  {author} {\bibinfo {author} {\bibfnamefont {J.~L.}\ \bibnamefont
  {Carini}}, \bibinfo {author} {\bibfnamefont {S.}~\bibnamefont {Kallush}},
  \bibinfo {author} {\bibfnamefont {R.}~\bibnamefont {Kosloff}}, \ and\
  \bibinfo {author} {\bibfnamefont {P.~L.}\ \bibnamefont {Gould}},\ }\href@noop
  {} {\bibfield  {journal} {\bibinfo  {journal} {Physical review letters}\
  }\textbf {\bibinfo {volume} {115}},\ \bibinfo {pages} {173003} (\bibinfo
  {year} {2015})}\BibitemShut {NoStop}%
\bibitem [{\citenamefont {Omiste}\ \emph {et~al.}(2018)\citenamefont {Omiste},
  \citenamefont {Flo\ss{}},\ and\ \citenamefont {Brumer}}]{2018-Coherent}%
  \BibitemOpen
  \bibfield  {author} {\bibinfo {author} {\bibfnamefont {J.~J.}\ \bibnamefont
  {Omiste}}, \bibinfo {author} {\bibfnamefont {J.}~\bibnamefont {Flo\ss{}}}, \
  and\ \bibinfo {author} {\bibfnamefont {P.}~\bibnamefont {Brumer}},\ }\href
  {\doibase 10.1103/PhysRevLett.121.163405} {\bibfield  {journal} {\bibinfo
  {journal} {Phys. Rev. Lett.}\ }\textbf {\bibinfo {volume} {121}},\ \bibinfo
  {pages} {163405} (\bibinfo {year} {2018})}\BibitemShut {NoStop}%
\bibitem [{\citenamefont {Ohayon}\ \emph {et~al.}(2018)\citenamefont {Ohayon},
  \citenamefont {Chocron}, \citenamefont {Hirsh}, \citenamefont {Glick-Magid},
  \citenamefont {Mishnayot}, \citenamefont {Mukul}, \citenamefont {Rahangdale},
  \citenamefont {Vaintraub}, \citenamefont {Heber}, \citenamefont {Gazit} \emph
  {et~al.}}]{2018-Weak}%
  \BibitemOpen
  \bibfield  {author} {\bibinfo {author} {\bibfnamefont {B.}~\bibnamefont
  {Ohayon}}, \bibinfo {author} {\bibfnamefont {J.}~\bibnamefont {Chocron}},
  \bibinfo {author} {\bibfnamefont {T.}~\bibnamefont {Hirsh}}, \bibinfo
  {author} {\bibfnamefont {A.}~\bibnamefont {Glick-Magid}}, \bibinfo {author}
  {\bibfnamefont {Y.}~\bibnamefont {Mishnayot}}, \bibinfo {author}
  {\bibfnamefont {I.}~\bibnamefont {Mukul}}, \bibinfo {author} {\bibfnamefont
  {H.}~\bibnamefont {Rahangdale}}, \bibinfo {author} {\bibfnamefont
  {S.}~\bibnamefont {Vaintraub}}, \bibinfo {author} {\bibfnamefont
  {O.}~\bibnamefont {Heber}}, \bibinfo {author} {\bibfnamefont
  {D.}~\bibnamefont {Gazit}},  \emph {et~al.},\ }\href@noop {} {\bibfield
  {journal} {\bibinfo  {journal} {Hyperfine Interactions}\ }\textbf {\bibinfo
  {volume} {239}},\ \bibinfo {pages} {57} (\bibinfo {year} {2018})}\BibitemShut
  {NoStop}%
\bibitem [{\citenamefont {Neynaber}\ and\ \citenamefont
  {Tang}(1980{\natexlab{a}})}]{1980-HeNe}%
  \BibitemOpen
  \bibfield  {author} {\bibinfo {author} {\bibfnamefont {R.}~\bibnamefont
  {Neynaber}}\ and\ \bibinfo {author} {\bibfnamefont {S.}~\bibnamefont
  {Tang}},\ }\href@noop {} {\bibfield  {journal} {\bibinfo  {journal} {The
  Journal of Chemical Physics}\ }\textbf {\bibinfo {volume} {72}},\ \bibinfo
  {pages} {5783} (\bibinfo {year} {1980}{\natexlab{a}})}\BibitemShut {NoStop}%
\bibitem [{\citenamefont {Hotop}\ \emph {et~al.}(1998)\citenamefont {Hotop},
  \citenamefont {Roth}, \citenamefont {Ruf},\ and\ \citenamefont
  {Yencha}}]{1998-Hotop}%
  \BibitemOpen
  \bibfield  {author} {\bibinfo {author} {\bibfnamefont {H.}~\bibnamefont
  {Hotop}}, \bibinfo {author} {\bibfnamefont {T.}~\bibnamefont {Roth}},
  \bibinfo {author} {\bibfnamefont {M.-W.}\ \bibnamefont {Ruf}}, \ and\
  \bibinfo {author} {\bibfnamefont {A.}~\bibnamefont {Yencha}},\ }\href@noop {}
  {\bibfield  {journal} {\bibinfo  {journal} {Theoretical Chemistry Accounts}\
  }\textbf {\bibinfo {volume} {100}},\ \bibinfo {pages} {36} (\bibinfo {year}
  {1998})}\BibitemShut {NoStop}%
\bibitem [{\citenamefont {Lorenzen}\ \emph {et~al.}(1986)\citenamefont
  {Lorenzen}, \citenamefont {Hotop},\ and\ \citenamefont {Ruf}}]{1986-alkali}%
  \BibitemOpen
  \bibfield  {author} {\bibinfo {author} {\bibfnamefont {J.}~\bibnamefont
  {Lorenzen}}, \bibinfo {author} {\bibfnamefont {H.}~\bibnamefont {Hotop}}, \
  and\ \bibinfo {author} {\bibfnamefont {M.~W.}\ \bibnamefont {Ruf}},\ }\href
  {\doibase 10.1007/BF01436681} {\bibfield  {journal} {\bibinfo  {journal}
  {Zeitschrift f{\"u}r Physik D Atoms, Molecules and Clusters}\ }\textbf
  {\bibinfo {volume} {1}},\ \bibinfo {pages} {261} (\bibinfo {year}
  {1986})}\BibitemShut {NoStop}%
\bibitem [{\citenamefont {Schohl}\ \emph {et~al.}(1990)\citenamefont {Schohl},
  \citenamefont {M{\"u}ller}, \citenamefont {Meijer}, \citenamefont {Ruf},
  \citenamefont {Hotop},\ and\ \citenamefont {Morgner}}]{1990-RgNa}%
  \BibitemOpen
  \bibfield  {author} {\bibinfo {author} {\bibfnamefont {S.}~\bibnamefont
  {Schohl}}, \bibinfo {author} {\bibfnamefont {M.~W.}\ \bibnamefont
  {M{\"u}ller}}, \bibinfo {author} {\bibfnamefont {H.~A.~J.}\ \bibnamefont
  {Meijer}}, \bibinfo {author} {\bibfnamefont {M.-W.}\ \bibnamefont {Ruf}},
  \bibinfo {author} {\bibfnamefont {H.}~\bibnamefont {Hotop}}, \ and\ \bibinfo
  {author} {\bibfnamefont {H.}~\bibnamefont {Morgner}},\ }\href {\doibase
  10.1007/BF01437526} {\bibfield  {journal} {\bibinfo  {journal} {Zeitschrift
  f{\"u}r Physik D Atoms, Molecules and Clusters}\ }\textbf {\bibinfo {volume}
  {16}},\ \bibinfo {pages} {237} (\bibinfo {year} {1990})}\BibitemShut
  {NoStop}%
\bibitem [{\citenamefont {Neynaber}\ and\ \citenamefont
  {Tang}(1980{\natexlab{b}})}]{1980-NeAr}%
  \BibitemOpen
  \bibfield  {author} {\bibinfo {author} {\bibfnamefont {R.}~\bibnamefont
  {Neynaber}}\ and\ \bibinfo {author} {\bibfnamefont {S.}~\bibnamefont
  {Tang}},\ }\href@noop {} {\bibfield  {journal} {\bibinfo  {journal} {The
  Journal of Chemical Physics}\ }\textbf {\bibinfo {volume} {72}},\ \bibinfo
  {pages} {6176} (\bibinfo {year} {1980}{\natexlab{b}})}\BibitemShut {NoStop}%
\bibitem [{\citenamefont {Doery}\ \emph
  {et~al.}(1998{\natexlab{c}})\citenamefont {Doery}, \citenamefont
  {Vredenbregt}, \citenamefont {Tempelaars}, \citenamefont {Beijerinck},\ and\
  \citenamefont {Verhaar}}]{1997-Verhhar}%
  \BibitemOpen
  \bibfield  {author} {\bibinfo {author} {\bibfnamefont {M.}~\bibnamefont
  {Doery}}, \bibinfo {author} {\bibfnamefont {E.}~\bibnamefont {Vredenbregt}},
  \bibinfo {author} {\bibfnamefont {J.}~\bibnamefont {Tempelaars}}, \bibinfo
  {author} {\bibfnamefont {H.}~\bibnamefont {Beijerinck}}, \ and\ \bibinfo
  {author} {\bibfnamefont {B.}~\bibnamefont {Verhaar}},\ }\href@noop {}
  {\bibfield  {journal} {\bibinfo  {journal} {Physical Review A}\ }\textbf
  {\bibinfo {volume} {57}},\ \bibinfo {pages} {3603} (\bibinfo {year}
  {1998}{\natexlab{c}})}\BibitemShut {NoStop}%
\bibitem [{\citenamefont {Gallagher}\ and\ \citenamefont
  {Pritchard}(1989{\natexlab{b}})}]{1989-GP}%
  \BibitemOpen
  \bibfield  {author} {\bibinfo {author} {\bibfnamefont {A.}~\bibnamefont
  {Gallagher}}\ and\ \bibinfo {author} {\bibfnamefont {D.~E.}\ \bibnamefont
  {Pritchard}},\ }\href@noop {} {\bibfield  {journal} {\bibinfo  {journal}
  {Physical review letters}\ }\textbf {\bibinfo {volume} {63}},\ \bibinfo
  {pages} {957} (\bibinfo {year} {1989}{\natexlab{b}})}\BibitemShut {NoStop}%
\bibitem [{\citenamefont {Oberheide}\ \emph
  {et~al.}(1997{\natexlab{b}})\citenamefont {Oberheide}, \citenamefont
  {Wilhelms},\ and\ \citenamefont {Zimmer}}]{1997-MCP}%
  \BibitemOpen
  \bibfield  {author} {\bibinfo {author} {\bibfnamefont {J.}~\bibnamefont
  {Oberheide}}, \bibinfo {author} {\bibfnamefont {P.}~\bibnamefont {Wilhelms}},
  \ and\ \bibinfo {author} {\bibfnamefont {M.}~\bibnamefont {Zimmer}},\
  }\href@noop {} {\bibfield  {journal} {\bibinfo  {journal} {Measurement
  Science and Technology}\ }\textbf {\bibinfo {volume} {8}},\ \bibinfo {pages}
  {351} (\bibinfo {year} {1997}{\natexlab{b}})}\BibitemShut {NoStop}%
\bibitem [{\citenamefont {Ohayon}\ \emph
  {et~al.}(2019{\natexlab{b}})\citenamefont {Ohayon}, \citenamefont
  {Rahangdale}, \citenamefont {Geddes}, \citenamefont {Berengut},\ and\
  \citenamefont {Ron}}]{2019-IS}%
  \BibitemOpen
  \bibfield  {author} {\bibinfo {author} {\bibfnamefont {B.}~\bibnamefont
  {Ohayon}}, \bibinfo {author} {\bibfnamefont {H.}~\bibnamefont {Rahangdale}},
  \bibinfo {author} {\bibfnamefont {A.~J.}\ \bibnamefont {Geddes}}, \bibinfo
  {author} {\bibfnamefont {J.~C.}\ \bibnamefont {Berengut}}, \ and\ \bibinfo
  {author} {\bibfnamefont {G.}~\bibnamefont {Ron}},\ }\href {\doibase
  10.1103/PhysRevA.99.042503} {\bibfield  {journal} {\bibinfo  {journal} {Phys.
  Rev. A}\ }\textbf {\bibinfo {volume} {99}},\ \bibinfo {pages} {042503}
  (\bibinfo {year} {2019}{\natexlab{b}})}\BibitemShut {NoStop}%
\bibitem [{\citenamefont {Ohayon}\ \emph
  {et~al.}(2015{\natexlab{b}})\citenamefont {Ohayon}, \citenamefont
  {W{\aa}hlin},\ and\ \citenamefont {Ron}}]{2015-Source}%
  \BibitemOpen
  \bibfield  {author} {\bibinfo {author} {\bibfnamefont {B.}~\bibnamefont
  {Ohayon}}, \bibinfo {author} {\bibfnamefont {E.}~\bibnamefont {W{\aa}hlin}},
  \ and\ \bibinfo {author} {\bibfnamefont {G.}~\bibnamefont {Ron}},\
  }\href@noop {} {\bibfield  {journal} {\bibinfo  {journal} {Journal of
  Instrumentation}\ }\textbf {\bibinfo {volume} {10}},\ \bibinfo {pages}
  {P03009} (\bibinfo {year} {2015}{\natexlab{b}})}\BibitemShut {NoStop}%
\bibitem [{\citenamefont {Whitaker}(2003{\natexlab{b}})}]{2003-Imaging}%
  \BibitemOpen
  \bibfield  {author} {\bibinfo {author} {\bibfnamefont {B.}~\bibnamefont
  {Whitaker}},\ }\href {https://books.google.co.il/books?id=m8AYdeM3aRYC}
  {\emph {\bibinfo {title} {Imaging in Molecular Dynamics: Technology and
  Applications}}}\ (\bibinfo  {publisher} {Cambridge University Press},\
  \bibinfo {year} {2003})\BibitemShut {NoStop}%
\end{thebibliography}%

\end{document}